\documentclass[showkeys,twocolumn,unsortedaddress,aps,prb,longbibliography]{revtex4-2}
\usepackage[utf8]{inputenc}
\usepackage[T1]{fontenc}
\usepackage{newtxtext,newtxmath}
\usepackage{graphicx}
\usepackage[colorlinks]{hyperref}
\usepackage{bm}
\usepackage{microtype}
\usepackage{physics}
\usepackage{xcolor}
\usepackage[caption=false]{subfig}
\usepackage{orcidlink}
\usepackage{relsize}

\newcommand{\xopr}[3][]{X_{#1}^{#2 #3}}

\begin{document}

\title{A method to treat strongly correlated topological superconductors in one and two dimensions}

\author{Kaushal K. Kesharpu\orcidlink{0000-0003-4933-6819}}
\email{kesharpu@theor.jinr.ru}
\affiliation{Bogoliubov Laboratory of Theoretical Physics, Joint Institute for Nuclear Research, Dubna, Moscow Region 141980, Russia}

\author{Evgenii A. Kochetov}
\email{kochetov@theor.jinr.ru}
\affiliation{Bogoliubov Laboratory of Theoretical Physics, Joint Institute for Nuclear Research, Dubna, Moscow Region 141980, Russia}

\author{Alvaro Ferraz}
\email{aferraz.iccmp@gmail.com}
\affiliation{International Institute of Physics - UFRN, Department of Experimental and Theoretical Physics - UFRN, Natal 59078-970, Brazil}

\date{\today}

\begin{abstract}
In the strong electron-electron (e-e) interaction limit each atomic site is constrained to be either empty or singly occupied. One can treat this scenario by fractionalizing the electrons into spin and charge degrees of freedom. We use the coherent state symbols associated with the lowest irreducible representation of the $su(2|1)$ superalgebra spanned by the Hubbard operators to solve the proposed models, as they implicitly take into account both the single particle occupation constraint and the fractionalization of the electrons. As an example, using the proposed method we solve the one dimensional Kitaev chain and two-dimensional BCS-Hubbard model to show the emergence of topological properties. The proposed procedure is quite general and can be used to analyze different lattice Hamiltonian, provided strong e-e correlation excludes doubly occupied states.
\end{abstract}

\keywords{Path Integral Method, Topological Superconductor, $su(2|1)$ Algebra, Strongly Correlated Electrons}

\maketitle

\section{introduction}
Topological quantum matters have been subjected to intense research due to their potential applications in transformational technologies; specifically, the possible application of the topological superconductors (SC) in quantum computers have been encouraging \cite{sarma-2015-major-zero}. Different types of one-dimensional (1D) topological SC (mainly heterostructures) have already been synthesized and experimentally verified \cite{flensberg-2021-engin-platf,frolov-2020-topol-super}. Their properties are usually explained through usual band theory (meaning e-e correlation has been neglected). However, if e-e correlation is non negligible, then the topological properties can not be analysed using the usual band theory; one needs to rely on other non-perturbative approaches \cite{avella2012-StronglyCorrelated,senechal2004-TheoreticalMethods,tsvelik2001-NewTheoretical}. In these approaches (under appropriate conditions) charge and spin degrees of freedom are treated separately (the so called fractionalization of spin and charge). Recently, there have been growing attempt to investigate these ``\emph{fractionalized}'' topological materials \cite{maciejko-2015-fract-topol-insul,irkhin_2022_TopologicalStates_JSupercondNovMagn,rachel-2018-inter-topol-insul,xie-2021-fract-chern,cai-2023-signat-fract}; especially, the strongly correlated topological SC using standard Abelian bosonization technique has been investigated intensively \cite{sagi_2017_FractionalChiral_PhysRevB,laubscher_2019_FractionalTopological_PhysRevRes,ronetti_2021_ClockModel_PhysRevB,delpozo_2023_FractionalTopology_PhysRevB}. As in 1D for the two-dimensions (2D) case, there also appears topological SC. In some of the 2D topological SC there exist nodes in the momentum space; consequentially they are called the nodal topological superconductors \cite{schnyder-2015-topol-surfac,schnyder-2012-types-topol,zhang-2019-higher-order,nayak-2021-eviden-topol,he-2018-magnet-field}. For the case of 2D Weyl quantum materials \cite{lu-2024-realiz-two,schnyder-2015-topol-surfac} in the momentum space an 1D fermi arc (also known as fermi string) connecting two Weyl nodes appears, which are protected as long as the Weyl nodes are present. As in the 1D case, in the 2D case also strong correlation plays important role in defining the topological properties \cite{schnyder-2015-topol-surfac,he-2018-magnet-field,nayak-2021-eviden-topol}.

The goal of this article is to provide a mathematically sound method to treat (in some defined limits) the topological superconducting systems in the presence of strong correlation. As an example, we apply the proposed method to the minimal 1D kitaev chain and 2D BCS-Hubbard model with strong e-e correlation, to show the emergence of the topologically non-trivial states. To this end, we first need to appropriately describe the strongly correlated system making no use of any mean-field approximation schemes. This can be achieved within the $su(2|1)$ path-integral technique that naturally incorporates the Gutzwiller projected electron operators which referred earlier on as Hubbard operators \cite{kesharpu-2025-reent-topol,kesharpu-2024-propos-realiz,kesharpu_2024_TopologicalHall_PhysRevB,kesharpu-2023-topol-hall,ferraz-2011-effec-action,ferraz-2023-connec-between}. Therefore extending our previous results we formulate a $su(2|1)$ path-integral approach to adequately treat the low energy limit of strongly correlated electron systems with emerging topological SC. It is well known that the conventional boson/fermion path-integral approaches \cite{senechal2004-TheoreticalMethods} adjusted to treat systems with strong e-e correlation contain a local constraint of no double occupancy (NDO) that gives rise to strongly coupled compact $U(1)$ lattice gauge theory that does not allow (even in 1D) for a reliable treatment in the presence of matter fields. In contrast, the $su(2|1)$ superalgebra path-integral representation is free from any such constraints.

An electron system is said to be strongly correlated if the leading energy scale in the problem is the on-site Coulomb repulsion energy $U$. In this case the low-energy sector of the underlying on-site Hilbert space should be
modified to exclude doubly occupied states. Such a modification results in an entirely new physics to account for. Formally strong correlations are naturally encoded into the projected electron (Hubbard) operators. They act directly in the restricted Hilbert space as opposed to the conventional electron operators which describe the unconstrained system. In contrast with the conventional fermion operators which generate  the standard  fermionic algebra, the new  operators obey more complicated commutation/anticommutation relations and are closed into a $su(2|1)$ superalgebra in its lowest representation \cite{wiegmann-1988-super-stron}. This representation acts on the physical on-site Hilbert space with three basis vectors $\ket{\uparrow},\ket{\downarrow},\ket{0}$;  and consists of nine operators
\begin{equation}
  \label{eq:su-2-1-operators}
  \xopr{\lambda}{\lambda'}=\ket{\lambda}\bra{\lambda'}, \text{  where } \ket{\lambda}, \ket{\lambda'}=\ket{\uparrow},\ket{\downarrow},\ket{0}.
\end{equation}
The resolution of unity in this space is
\begin{equation}
\hat{I}= \ket{0}\bra{0} + \ket{\uparrow}\bra{\uparrow} + \ket{\downarrow}\bra{\downarrow} = \sum\limits_{\lambda} \xopr{\lambda}{\lambda}.
\label{space}\end{equation}
Using $\hat{I}$ we can single out eight independent generators of $su(2|1)$ algebra. The special important property of this representation is the contraction of generators:
\begin{equation}
  \xopr{\lambda}{\lambda'} \xopr{\lambda''}{\lambda'''}= \delta_{\lambda' \lambda''}\xopr{\lambda}{\lambda'''}.
\end{equation}
In other words the condition of either a singly occupied or an empty state is automatically satisfied.

The main results and structure of the article is as follows. Sec. \ref{sec:su21-coherent-state} and \ref{sec:su21-path-integral} contain the theoretical part of the  problem. There we introduce the coherent state (CS) symbols of the Hubbard operator associated with the $su(2|1)$ superalgebra, and thereafter solve the Hubbard Hamiltonian given in terms of CS Hubbard operators using $su(2|1)$ path integral method. In Sec. \ref{sec:topol-superc} and \ref{eq:kx-ky-mu-less} the application of the method to the 1D and 2D case is shown. In the Sec. \ref{sec:conclusion} the physical realization and applicability of the model have been discussed.

\section{$su(2|1)$ coherent-state dynamics}
\label{sec:su21-coherent-state}

The $su(2|1)$ superalgebra can be thought of as the simplest possible extension of the conventional spin $su(2)$ algebra to incorporate fermionic degrees of freedom \cite{ferraz-2011-effec-action}. Namely, the  bosonic sector of the $su(2|1)$ consists of three bosonic superspin operators,
\begin{equation}
Q^{+}=X^{\uparrow\downarrow},\quad Q^{-}=X^{\downarrow\uparrow},\quad
Q^{z}=\frac{1}{2}(X^{\uparrow\uparrow}-X^{\downarrow\downarrow})
\label{eq:bosonic-sector}
\end{equation}
closed into $su(2)$, and a bosonic operator $X^{00}$ that generates a $u(1)$ factor of the maximal even subalgebra $su(2)\times u(1)$ of $su(2|1)$. The fermionic sector is constructed out of four operators
\begin{equation}
  \label{eq:fermionc-sector}
  X^{\uparrow 0},\quad X^{\downarrow 0}, \quad X^{0 \uparrow}, \quad X^{0 \downarrow},
\end{equation}
that transform in a spinor representation of $su(2)$.

The normalizable CS associated with the lowest irreducible representation of the $su(2|1)$ superalgebra spanned by Hubbard operators take the form
\begin{equation}
  \label{eq:cs-symbol-su-1}
  \begin{aligned}
    \ket{z, \xi}&= \left[\frac{\exp \left( z \xopr{\downarrow}{\uparrow} + \xi \xopr {0}{ \uparrow}\right)}{\sqrt{ 1 + \bar{z}z + \bar{\xi} \xi}}  \right] \ket{\uparrow} = \frac{ \ket{\uparrow} + z \ket{\downarrow} + \xi \ket{0}}{\sqrt{ 1 + \bar{z}z + \bar{\xi} \xi}}.
  \end{aligned}
\end{equation}
Where $z$ is a complex parameter (representing bosonic field) and $\xi$ is an odd complex Grassmann parameter (representing fermionic field) \cite{ferraz-2011-effec-action}. The $z$ and $\xi$ characterize the inhomogeneous (proper) coordinates of a point on a supersphere, i.e.
$$(z,\xi) \in S^{2|2}\simeq CP^{1|1}=SU(2|1)/U(1|1).$$
 Here $CP^{1|1}$ stands for a complex projective superspace with a complex dimension $(1,1)$. It can be thought of as the minimal superextension of an ordinary projective space $CP^1$ homeomorphic to a two-sphere, $CP^1\simeq S^2$. The $CP^{1|1}$ manifold serves as a classical phase space for the Hubbard operators. The odd Grassmann parameter ($\xi$) appears in Eq. (\ref{eq:cs-symbol-su-1}) due to the fact that $X^{\downarrow 0}$ is a fermionic operator in contrast with the bosonic operator $X^{\downarrow\uparrow}$. The product $\xi X^{0\uparrow}$ represents therefore a bosonic quantity as required.

 One can use as well the homogeneous coordinates $(z^1,z^2,\theta)$ on the supersphere. In this case the aforementioned inhomogeneous coordinates of the supersphere ($z$, $\xi$) are represented as:
 \begin{equation}
   \label{eq:inhom-in-terms-of-homo}
    z=z^1/z^2, \: \xi=\theta/z^2,\quad \text{with } z^2\neq 0.
 \end{equation}
 Then the coset $CP^{1|1}$ manifold is defined by the condition
 $$\left| z^1 \right|^2+ \left|z^2  \right|^2+\bar\theta\theta=1.$$
 The $SU(2|1)$ supergroup acts on $CP^{1|1}$ according to
$$Z\to Z^g=gZ,$$ where $$Z=\left(z^1,z^2,\theta \right)^T$$ and $g\in SU(2|1).$ This generates a corresponding transformation of the inhomogeneous coordinates, $(z,\xi)$. In particular, if one chooses $g$ to represent a pure spin rotation,
$$g=
 \left(\begin{array}{lll}
u & v&0 \\
-\overline{v} & \overline{u}&0\\
0&0&1
\end{array}
\right),\quad
 \left(\begin{array}{ll}
u & v \\
-\overline{v} & \overline{u}%
\end{array}
\right) \in \mathrm{SU(2)},
$$
one gets
\begin{equation}
z\to \frac{uz+v}{-\overline{v}z+\overline{u}},\quad \xi\to \frac{\xi}{-\overline{v}z+\overline{u}}.
\label{2field}\end{equation}
Note that both the bosonic ($z$) and fermionic ($\xi$) fields transform themselves under $SU(2)$ spin rotations.

For $\xi =0$ the $su(2|1)$ CS in Eq. (\ref{eq:cs-symbol-su-1}) reduces to the ordinary spin $su(2)$ CS,
\begin{equation}
  \label{eq:cs-symbol-xi-0}
  \begin{aligned}
  \ket{z, \xi=0} &= \ket{z}_{s=1/2} \equiv \ket{z},
  \end{aligned}
\end{equation}
where
\begin{equation}
  \label{eq:defn-cs-symb-xi-0}
  \begin{aligned}
    \ket{z}= \left[\frac{\exp \left( z S^{-} \right)}{\sqrt{1+\left| z \right|^{2}}}  \right] \ket{\uparrow} = \frac{ \ket{\uparrow} + z \ket{\downarrow}}{\sqrt{1+\left| z \right|^{2}}}.
  \end{aligned}
\end{equation}
Here $S^{-}$ is the $su(2)$ spin ladder operator, which is analogous to the $su(2|1)$ operator $Q^{-}$ in Eq. (\ref{eq:bosonic-sector}). The complex number $z$ is now a stereographic coordinate of a point on an ordinary sphere (instead of the supersphere as in $su(2|1)$ case) $$z\in S^2\simeq CP^1=SU(2)/U(1).$$ The spin operators $\vec S$ obey the standard commutation relations
\begin{equation}
  [S^{z},S^{\pm}]=\pm S^{\pm},\quad [S^{+},S^{-}]=2S^{z}, \quad \vec S^2=3/4.
  \label{1.2b}
\end{equation}
These operators coincide with bosonic generators $\vec Q,$ of $su(2|1)$ at half filling, in which case the on-site Hilbert space is reduced and spanned only by the vectors
$|\uparrow\rangle, \,|\downarrow\rangle$. When $z=0$ the CS in Eq. (\ref{eq:cs-symbol-su-1}) $|\xi\rangle\equiv |z=0,\xi\rangle$ represents a pure fermionic CS; in the sense either empty of singly occupied state is possible (it is the same as two fermionic particles can not occupy the same state).

At this stage it is helpful to introduce the important notion of the covariant (Berezin) symbol for
a Hubbard $X$ operators. It can also be referred to as a coherent-state symbol and it is defined as follows
\begin{equation}
X_{\text{cov}}:=\langle z,\xi|X|z,\xi\rangle.
\label{symbol}\end{equation}
Explicitly, the fermionic $su(2|1)$ operators from Eq. (\ref{eq:fermionc-sector}) reads
\begin{equation}
  \label{eq:fermionic-su-2-1}
  \begin{aligned}
    &X^{0\downarrow}_{\text{cov}} = -\frac{z\bar\xi}{1+|z|^2},\quad
    &&X^{\downarrow 0}_{\text{cov}}=-\frac{\bar z \xi}{1+|z|^2},\\
    &X^{0\uparrow}_{\text{cov}}=-\frac{\bar\xi}{1+|z|^2}, \quad
    &&X^{\uparrow 0}_{\text{cov}}=-\frac{\xi}{1+|z|^2}.
  \end{aligned}
\end{equation}
Similarly the bosonic $su(2|1)$ operators from Eq. (\ref{eq:bosonic-sector}) read
\begin{equation}
  \begin{aligned}
    \label{eq:bosonic-su-2-1}
    &Q^{+}_{\text{cov}}=S^{+}_{\text{cov}}\left(1-X^{00}_{\text{cov}}\right),\quad Q^{-}_{\text{cov}}=S^{-}_{\text{cov}}\left(1-X^{00}_{\text{cov}}\right),\\
    &Q^z_{\text{cov}}=S^z_{\text{cov}}\left(1-X^{00}_{\text{cov}}\right).
  \end{aligned}               
\end{equation}
Here the covariant symbol of the hole number operator reads
\begin{equation}
X_{\text{cov}}^{00}=\frac{\bar\xi\xi}{1+|z|^2}.
\label{1.3a}
\end{equation}
The CS symbols of the $su(2)$ generators used in Eq. (\ref{eq:bosonic-su-2-1}) are evaluated as $S_{\text{cov}}:=\langle z|S|z\rangle$, and explicitly represented as:
\begin{equation}
  \label{eq:su-2-generator-cs-symb}
  \begin{aligned}
    &S^{+}_{\text{cov}}=\frac{z}{1+|z|^2}, \quad
    S^{-}_{\text{cov}}=\frac{\bar z}{1+|z|^2},\\
    &S^z_{\text{cov}}=\frac{1}{2}\left(\frac{1-|z|^2}{1+|z|^2}\right).
  \end{aligned}    
\end{equation}
For a compact simple (super)algebra these symbols are in one-to-one correspondence with the algebra generators. This is the case for both the $su(2|1)$ and $su(2)$ algebras. 

\section{$su(2|1)$ path integral}
\label{sec:su21-path-integral}

Given the Hamiltonian $H$ in terms of the Hubbard operators, the corresponding imaginary time phase-space action takes on the form,
\begin{equation}
  \label{eq:action-phase-space}
  \begin{aligned}
    S=-\int\limits_0^{\beta} \left\langle z, \xi \left| \frac{\partial}{\partial t} + H \right|z,\xi \right\rangle dt,
  \end{aligned}
\end{equation}
where the $su(2|1)$ symplectic potential is defined as
\begin{equation}
  \label{eq:sympletic-pot}
  \begin{aligned}
    \left\langle z,\xi \left| - \frac{\partial}{\partial t} \right| z, \xi \right\rangle = \frac{1}{2} \left( \frac{\dot{\bar{z}} z - \bar{z} \dot{z} + \dot{\bar{\xi}} \xi - \bar{\xi} \dot{\xi}}{1+\left| z \right|^{2}+\bar{\xi}} \right).
  \end{aligned}
\end{equation}
The $H$ in Eq. (\ref{eq:action-phase-space}) can represent arbitrary Hamiltonian (provided it is valid only on the constrained Hilbert space). However, as an example, here we consider one of the simplest $U=\infty$ Hubbard model:
\begin{equation}
  \label{3.10}
  H=-t\sum_{\left\langle i,j \right\rangle,\sigma}X^{\sigma 0}_{i}X^{0\sigma}_j + \text{H.c.} + \mu\sum_iX^{00}_i.
\end{equation}
The chemical potential term is added to fix the total number of vacancies. The CS symbol of $H$ becomes
\begin{equation}
H_{\text{cov}}=-t\sum_{\left\langle i,j \right\rangle,\sigma}X^{\sigma 0}_{i, cov}X^{0\sigma}_{j, cov} + \text{H.c.} + \mu\sum_iX^{00}_{i, cov}.
  \label{3.10a}
\end{equation}
The partition function takes a form of the $su(2|1)$ CS path integral \cite{ferraz-2011-effec-action}:
\begin{equation}
  \label{eq:part-function}
Z=\int D\mu_{su(2|1)} (z,\xi ) \exp S.
\end{equation}
Where the $su(2|1)$ invariant measure reads
\begin{equation}
  \label{eq:measure-su-2-1}
  \begin{aligned}
    D\mu_{su(2|1)}\left( z, \xi \right)=\prod_{i,t}  \frac{d\bar{z}_{i}(t) \: dz_{i}(t) \: d\bar{\xi}_{i}(t) \: d\xi_{i}(t)}{ 1 + \left| z_{i}(t) \right|^{2} + \bar{\xi}_{i}(t) \xi_{i}(t)}.
  \end{aligned}
\end{equation}

Representation Eq. (\ref{eq:part-function}) exhibits a global $SU(2)$ invariance. This is a direct consequence of the fact that the Hubbard operators $X^{\sigma 0}, X^{0\sigma}$ transform in the fundamental representation of the $su(2)$ algebra. The Eq. (\ref{eq:part-function}) is also time-reversal ($\mathcal{T}$) invariant. Under $\mathcal{T}$ the phase-space manifold transforms to
\begin{equation}
z_i(t) \to -1/\bar{z}_i \left( \beta -t \right) \quad \xi_i (t)\to \bar{\xi}_i \left( \beta -t \right) /\bar{z}_i \left( \beta - t \right).
\label{eq:TR-invariant}
\end{equation}
Using Eq. (\ref{eq:TR-invariant}) in Eq. (\ref{eq:fermionic-su-2-1}) we get the corresponding transformation of the $X$ operators. Substituting these in Eq. (\ref{3.10}) we will see the $\mathcal{T}$ invariant of the Hamiltonian.


To proceed further it is helpful to make a change of variables
\begin{equation}
z\to z,\quad \xi\to\xi\sqrt{1+|z|^2}
\label{change}\end{equation}
to bring the measure into a more tractable form
\begin{equation}
  \label{changem}
  \begin{aligned}
    &D\mu_{su(2|1)}
    = \frac{d\bar zdzd\bar\xi d\xi}{2\pi i(1+\bar zz+\bar\xi\xi)} \to
      \frac{d\bar zdzd\bar\xi d\xi}{2\pi i(1+\bar zz)^2}(1-\bar\xi\xi),
  \end{aligned}    
\end{equation}
or in other words
\begin{equation}
  \label{changem2}
  \begin{aligned}
    \frac{d\bar zdzd\bar\xi d\xi}{2\pi i(1+\bar zz)^2}(1-\bar\xi\xi)
    = D\mu_{su(2)}D\mu_{u(1)}(1-\bar\xi\xi).
  \end{aligned}    
\end{equation}
Here $D\mu_{su(2)}=\frac{d\bar zdz}{2\pi i(1+\bar zz)^2}$ stands for the spin $SU(2)$ invariant measure, and $D\mu_{u(1)}=d\bar\xi d\xi$ denotes a standard $U(1)$ invariant fermionic measure. The full measure Eq. (\ref{changem}) differs from the standard product of the spin and fermionic measures $D\mu_{su(2)}\,D\mu_{u(1)}$ by a factor of $(1-\bar\xi\xi)$.

The on-site $su(2|1)$ partition function at $H=0$ is equal to a dimensionality of the underlying on-site Hilbert space for strongly correlated electrons \cite{ferraz-2011-effec-action}:
\begin{equation}
  \label{eq:z-u-infty}
Z_{U=\infty}^{(0)}= \text{Tr} \: \mathrm{e}^{-\beta\mu X^{00}}\mid_{\mu=0}= \int D\mu_{su (2|1)} \: \exp S =3.
\end{equation}
This is due to the fact that our space is composed of an electron spin-up, a spin-down and a vacant state. For this case the measure is given by Eq. (\ref{changem}) and the on-site action reads
\begin{eqnarray}
S=\int_0^{\beta} \left[ \imath a^{(0)}- \bar\xi \left(\partial_{t} + \imath a^{(0)}\right)\xi   \right] dt.
\label{e2}\end{eqnarray}
The action involves the time component of the connection one-form of the spin (magnetic monopole) bundle \cite{stone-1989-super-quant},
\begin{equation}
  \label{eq:connection-form}
  \imath a^{(0)}= - \left\langle z \left| \partial_{\tau} \right| z \right\rangle = \frac{1}{2} \left( \frac{\dot{\bar{z}} z - \bar{z} \dot{z}}{1+ \left| z \right|^{2}} \right).
\end{equation}
Here $\dot{z}$ is the time derivative. If the extra factor in the measure of Eq. (\ref{changem}) is dropped out, one gets instead $Z_{U=\infty}=2 \times 2$, which is the dimensionality of the spin space (spin up and spin down states) $\times$  the  dimensionality of the spinless fermionic space (empty state and spinless one-fermion state).

Having said that, it will be instructive to compute the reduced partition function $\tilde{Z}$ found by dropping $(1-\bar{\xi} \xi)$ in the measure. Explicitly, the reduced partition function reads
\begin{equation}
\tilde{Z}=
\int D\mu_{su(2)} \: D\mu_{u(1)} \exp S
\label{55.1}.
\end{equation}
The action $S$ is defined as
\begin{equation}
  \label{eq:action-cov}
  \begin{aligned}
    S = &\sum\limits_{i} \int\limits_{0}^{\beta} \left[ \imath a_{i}^{(0)} - \bar{\xi}_{i} \left( \partial_{t} + \imath a_{i}^{(0)} \right) \xi_{i} - H_{\text{cov}} \right] \: dt,
  \end{aligned}
\end{equation}
where
\begin{eqnarray}
H_{\text{cov}}=-t\sum_{ij}\bar\xi_i\xi_j e^{ia_{ji}} + \mathrm{H.c.} + \mu \sum\limits_{i} \bar{\xi}_{i}\xi_{i}. \label{5.3}
\end{eqnarray}
$H_{\text{cov}}$ is found by applying the representations Eq. (\ref{eq:fermionic-su-2-1}) to Eq. (\ref{3.10a}). In Eq. (\ref{5.3}) the $a_{ij}=-i\log \langle z_i|z_j\rangle$ is a lattice spatial component of the spin connection that is expressed through $su(2)$ coherent states Eq. (\ref{eq:defn-cs-symb-xi-0}) \cite{stone-1989-super-quant}.


It can be shown that the partition functions $Z$ and $\tilde Z$ in 1D capture the same low energy physics as in $U=\infty$ model. In a 1D setup the $U=\infty$ Hubbard model is known to be exactly solvable. The exact ground-state energy can be recovered within the path-integral representation Eq. (\ref{eq:part-function}), i.e.
\begin{equation}
  \label{eq:path-inegral-1D}
  Z_{U=\infty}=\prod_{k:\,T_k<0}e^{-\beta T_k}(1+ \mathcal{O} (1)); \quad \mu\beta\to\infty.
\end{equation}
Here $T_k \equiv -2t\cos k -\mu$; $k\in$ BZ. The ground-state energy density then becomes
\begin{equation}
E_{gr}=-\frac{2t}{\pi}\sin(\pi\delta);\quad \delta=\frac{1}{\pi}\arccos(\frac{-\mu}{2t}), \quad t\ge 0.
\label{e3}\end{equation}
The result coincides with the exact 1D result for the $U=\infty$ Hubbard model \cite{ogata-1990-bethe-ansat}. The same result follows from the reduced spin-charge path-integral representation Eq. (\ref{55.1}).

A geometric interpretation of the action is important since it allows us to address the underlying symmetry properties in a direct way. For instance, under a {\it global} $SU(2)$ rotation,
\begin{eqnarray} z_i\rightarrow
\frac{uz_i+v}{-\overline{v}z_i+\overline{u}}.
\label{tr}
\end{eqnarray}
We get
\begin{equation}a_i^{(0)}\to a^{(0)}_i-\partial_t\theta_i,\quad
a_{ij}\rightarrow a_{ij}+\theta_j-\theta_i,
\label{5.4}
\end{equation}
where
\begin{eqnarray}
\theta_i=-\frac{i}{2}\log \frac{-v\overline{z}_i+u}{-\overline{v}z_i+\overline{
u}}, \quad \left(\begin{array}{ll}
u & v \\
-\overline{v} & \overline{u}%
\end{array}
\right) \in \mathrm{SU(2)}.
\end{eqnarray}
The action Eq. (\ref{eq:action-cov}) is globally SU(2) invariant as described before. The above transformations emerge as conventional  transformations of the lattice connections in a vector bundle under a change of local frame \cite{nakahara2018-GeometryTopology}. Such a framework also allows us to directly address the topological properties of the electron system. For instance, the $U(1)$ Bloch  bundle  connections determine topological invariants that characterize the topologically non-trivial electron states in the anomalous Hall effect \cite{kesharpu-2023-topol-hall}.

Fractionalization of the lattice electrons into spinless fermions and localized lattice spins is believed to be a hallmark of strong correlation incorporated in this path integral Eq. (\ref{eq:part-function}). However, there is still no route available to compute that integral except in some trivial cases. This can be traced to the observation that there is a significant entanglement between the  bosonic and fermionic degrees of freedom through the path-integral $su(2|1)$ measure Eq. (\ref{eq:measure-su-2-1}). In contrast, the reduced measure --- found by dropping $\left( 1 - \bar{\xi} \xi \right)$ in Eq. (\ref{changem2}) --- splits into two independent pieces; the $SU(2)$ spin measure $D\mu_{su(2)}$ and the $U(1)$ charge measure $D\mu_{u(1)}$.

Our conjecture  is then that the alternative simplified spin-charge path-integral Eq. (\ref{55.1}), under special conditions, can be used in place of Eq. (\ref{eq:part-function}) to treat strongly correlated systems. It properly encodes strong correlation as  the action in Eq. (\ref{eq:action-cov}) is solely expressed in terms of the spinless fermions and the localized $su(2)$ spins. As a result, double electron occupancy continues to be prohibited.

Such a simplification works in the case where elementary charge and spin excitations are separated by a large energy gap. This is the main reason why the partition function $Z$ be replaced by the reduced partition function $\tilde{Z}$ in the low energy limit. The last one is clearly more easier to deal with. Such a conjecture is based on the fact that the charge-spin mixing is not very important in the case where the spin and the charge energy scales are essentially different from each other. This is believed to be the case for strongly correlated lattice electrons at small doping. In this case the hopping energy is of order $t$  whereas the nn spin exchange energy $J\sim t^2/U \ll t$.

\section{Application to the 1D Kitaev chain}
\label{sec:topol-superc}
\begin{figure*}[tb]
  \centering
  \subfloat[][]{\includegraphics[width=0.9\textwidth]{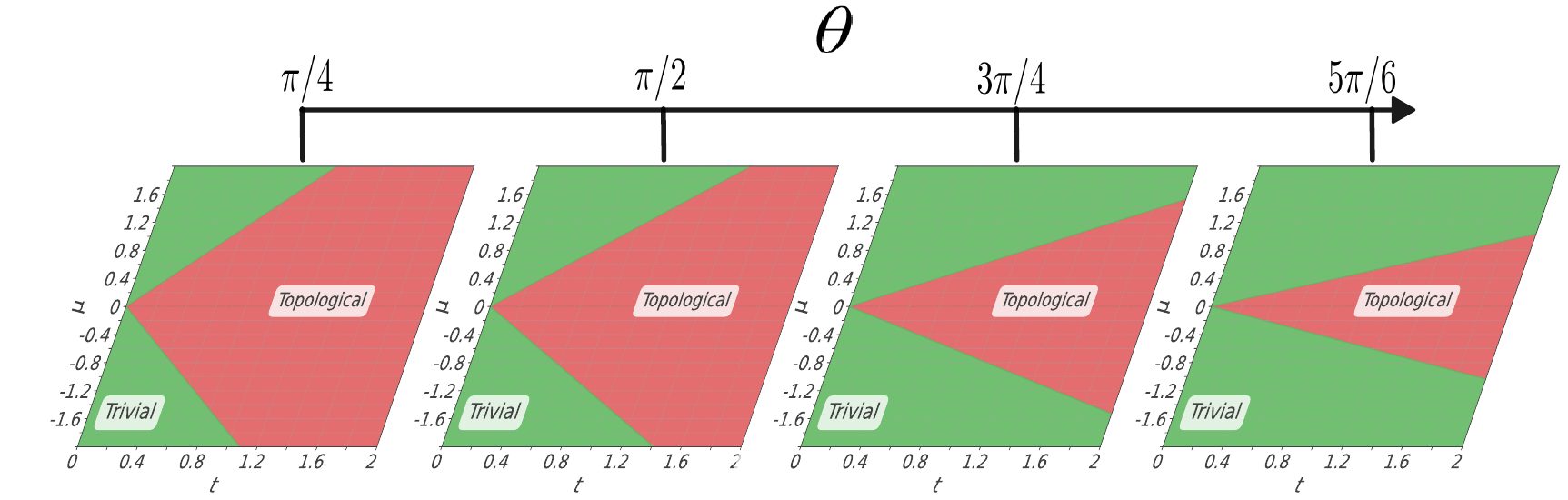}}
  \caption{Topological phase diagram dependence on $\theta$. The topological non-trivial phase (red) occurs when the condition, Eq. (\ref{eq:topo-phase-spiral}) is satisfied. With increase in $\theta$ the topologically trivial phase (green) increases. The non-trivial phase is absent for $\theta=\pi$ (anti-ferromagnetic).}
  \label{fig:polar-phase-plot}
\end{figure*}

The effective $su(2|1)$ path-integral representation of the partition function is convenient in investigating the properties of the 1D topological SC nanowires (NW) with strong e-e correlation. For similar systems without e-e correlation (band theory is applicable) \cite{oreg-2010-helic-liquid,lutchyn-2018-major-zero} the external magnetic field is a necessary ingredient to project out the upper band (remove the degeneracy at $k$=0; $k$ being the crystal momentum). In the lower band there remains only effective ``\emph{chiral spinless}" fermions \cite{streda-2003-antis-spin} --- the necessary ingredient for topological SC to emerge. In fact, a large magnetic field is desirable for opening a large gap at $k=0$. However, a large magnetic field also destroys SC. So there are opposite requirements to be fulfilled. 1D topological SC NW with strong e-e correlation provides a way to circumvent this situation, as the degeneracy is lifted at $k=0$ by default (the NDO constraint). However, one should not forget that the spin degree of freedom is still present (spins are not paralysed along single direction); this allows for imposing different spin structure and analyse the Hamiltonian. The path-integral representation of the partition function Eq. (\ref{55.1}) offers a procedure to treat these systems.

The Kitaev chain is the archetypal model for topological SC \cite{kitaev-2001-unpair-major,alicea-2012-new-direc}. We assume a NW with strong e-e correlation is placed over an SC substrate. Superconductivity in the NW is induced due to its proximity to the SC substrate. The SC order parameter $\Delta$ can be either $s$-wave or extended $s$-wave. If $\Delta$ is isotropic in momentum ($k$) space then it is called extended $s$-wave. We assume an externally applied magnetic field perpendicular to the NW axis; direction of the magnetic field is taken as $z$-axis.

To model this system we start from the Hamiltonian of the usual NW without e-e correlation \cite{oreg-2010-helic-liquid}, and later add correlations to find its effective $su(2|1)$ representation. The non-interacting Hamiltonian reads
\begin{equation}
  \label{eq:ham-1d-NW-weak-corr}
  \begin{aligned}
    H=& -t \sum\limits_{i,\sigma} c^{\dagger}_{i,\sigma}c_{i+1,\sigma} - \Delta \sum\limits_{i,\sigma} c^{\dagger}_{i,\sigma} c^{\dagger}_{i+1,\sigma'} + \text{H.c.}\\
    & \quad +\sum\limits_{i,\sigma} \left(B_{z} \sigma_{z}  \right) \: c_{i,\sigma}^{\dagger}c_{i,\sigma} + \mu \sum\limits_{i,\sigma}c_{i,\sigma}^{\dagger}c_{i,\sigma}.
  \end{aligned}
\end{equation}
Here, $t$ is the hopping parameter; $\Delta$ is the $s$-wave SC order parameter; $\mu$ is the chemical potential; $c_{i,\sigma}$ ($c_{i,\sigma}^{\dagger}$) is the electron annihilation (creation) operator at the $i$-th site with spin $\sigma$; $B_{z}$ is the externally applied magnetic field strength; $\sigma_{z}$ is the Pauli matrix. The e-e correlation is added by extending the Hamiltonian Eq. (\ref{eq:ham-1d-NW-weak-corr}) to include onsite Coulomb repulsion ($U$):
$$H \to  H_{U} \equiv H + U \sum\limits_{i} n_{i,\sigma}n_{i,\sigma'}.$$ Here, $n_{i,\sigma}=c_{i,\sigma}^{\dagger}c_{i,\sigma}$ is the number operator which counts the number of electron with spin $\sigma$ at the $i$-th site. This Hamiltonian under strong correlation, i.e. $U \to \infty$ reads
\begin{equation}
  \label{eq:ham-1d-wire}
  \begin{aligned}
    &H = -t \sum\limits_{i, \sigma} X_{i}^{\sigma 0}X_{i+1}^{0 \sigma} 
                     + \Delta \sum\limits_{i} \left( X_{i}^{\sigma 0} X_{i+1}^{\sigma' 0} -  X_{i}^{\sigma' 0} X_{i+1}^{\sigma 0}   \right) + \text{H.c.}\\
                    &\qquad \qquad + B_{z} \sum\limits_{i} \left( \xopr[i]{\sigma}{0} \xopr[i]{0}{\sigma} - \xopr[i]{\sigma'}{0} \xopr[i]{0}{\sigma'}   \right) + \mu \sum\limits_{i} X_{i}^{00}.
  \end{aligned}
\end{equation}
Where the Hubbard operators $X^{pq}$ through conventional fermionic operators are defined as
\begin{equation}
  \label{eq:Hubbard-oprs}
  \begin{aligned}
    &X_{i}^{0\sigma} = c_{i\sigma} (1-n_{i \sigma'}),
      \quad X_{i}^{\sigma 0} =(1-n_{i \sigma'})c_{i\sigma}^{\dagger},\\
    &X_{i}^{0 0} = X_{i}^{0\sigma} X_{i}^{\sigma 0} = c_{i\sigma} (1-n_{i \sigma'}) (1-n_{i \sigma'})c_{i\sigma}^{\dagger}.
  \end{aligned}
\end{equation}
After this the Hamiltonian can be represented in terms $su(2|1)$ spinon bosonic fields $z$ and holon fermionic fields $\xi$ using Eqs. (\ref{eq:fermionic-su-2-1}) -- (\ref{eq:su-2-generator-cs-symb}):
\begin{equation}
  \label{eq:effectiv-ham-trans}
  \begin{aligned}
    H =&- t \sum\limits_{i} \xi_{i+1} \bar{\xi}_{i} \: a_{i}  - \Delta\sum\limits_{i}\bar{\xi}_{i+1}\bar{\xi}_{i}\: \Delta_{i}+ \text{H.c.}\\
       &\qquad + \sum\limits_{i} \xi_{i} \bar{\xi}_{i} \left(\mu + B_{z}\beta_{i}\right),
  \end{aligned}
\end{equation}
where,
\small
\begin{equation}
  \label{eq:def-aij}
  \begin{aligned}
    &a_{i} \equiv \frac{1 + \bar{z}_{i+1} z_{i}}{\sqrt{\left( 1 + \left| z_{i+1} \right|^{2}\right)\left( 1 + \left| z_{i} \right|^{2}\right)}},\quad \beta_{i} \equiv \frac{1 - \left| z_{i}^{2} \right|}{1 + \left| z_{i}^{2} \right|}\\
    &\Delta_{i} \equiv \frac{z_{i} - z_{i+1}}{\sqrt{\left( 1 + \left| z_{i} \right|^{2}\right) \left( 1 + \left| z_{i+1} \right|^{2}\right)}}.
  \end{aligned}
\end{equation}  
\normalsize
are the spatially dependent function of the bosonic operator $z_{i}$; using $su(2)$ generators, Eq. (\ref{eq:su-2-generator-cs-symb}), one can represent the terms in Eq. (\ref{eq:def-aij}) more familiar $SU(2)$ operators.

In the second quantization language $\xi_{i}$ ($\bar{\xi}_{i}$) annihilates (creates) a spinless fermion at the $i$-th site. Similarly, $z_{i}$ ($\bar{z}_{i}$) annihilates (creates) a spinfull boson at the $i$-th site. In Eq. (\ref{eq:effectiv-ham-trans}) the $\xi$ and $z$ emerge as low-energy degrees of freedom; this allows to treat them on the same footing. $a_{i}$, $\Delta_{i}$ and $\beta_{i}$ are functions only of the spin degrees of freedom; their functional forms are different for different of the spin structure.

First we analyse the topological properties due to presence of spiral spin structures in the NW. It is defined as:
\begin{equation}
  \label{eq:spin-config-spiral-xy}
  \vec{S}_{i} = \left( S_{i}^{x}, S_{i}^{y}, S_{i}^{z} \right) = \frac{1}{2}\left(\cos \theta_{i}, \sin \theta_{i}, 0  \right).
\end{equation}
$\theta_{i}=\vec{q} \cdot \vec{r}_{i}$ gives the spatial dependence of the spin; $\vec{q}$ is the spin modulation vector (for 1D both $\vec{q}$ and $\vec{r}_{i}$ should be scalars). Using Eq. (\ref{eq:spin-config-spiral-xy}) in Eq. (\ref{eq:su-2-generator-cs-symb}) one can represent $$z_{i}= \mathrm{e}^{i \theta}.$$ Using this $z_{i}$ in Eq. (\ref{eq:def-aij}) we find
\begin{equation}
  \label{eq:spiral-aij}
  \begin{aligned}
    a_{i} = \mathrm{e}^{- \imath \theta/2}\cos \left( \theta/2 \right), \: \Delta_{i} = - \imath \sin \left( \theta/2 \right) \mathrm{e}^{\imath \theta_{i}}, \: \beta_{i}=0.
  \end{aligned}
\end{equation}
Here we have used the fact that $\left(\theta_{i+1}-\theta_{i}  \right)= \vec{q} \cdot \left( \vec{r}_{i+1} - \vec{r}_{i}\right)\equiv \theta$. Substituting Eq. (\ref{eq:spiral-aij}) in the Hamiltonian, Eq. (\ref{eq:effectiv-ham-trans}), and making the gauge transformation $\xi_{i} \to \xi_{i}\mathrm{e}^{\imath (\theta_{i}/2-\pi/4)}$ we get
\small
\begin{equation}
  \label{eq:effectiv-ham-spiral-spin}
  \begin{aligned}
    H=-t \cos \frac{\theta}{2}\sum\limits_{i} \xi_{i+1} \bar{\xi}_{i} - \Delta \sin \frac{\theta}{2}\sum\limits_{i}\bar{\xi}_{i+1}\bar{\xi}_{i} + \text{H.c.} + \mu\sum\limits_{i} \xi_{i}\bar{\xi}_{i}.
  \end{aligned}
\end{equation}
\normalsize
In Eq. (\ref{eq:effectiv-ham-spiral-spin}) it is interesting to see that the effect of magnetic field is completely absent (as $\beta_{i}=0$).

The Hamiltonian, Eq. (\ref{eq:effectiv-ham-spiral-spin}), is analogous to the Kitaev spin chain \cite{kitaev-2001-unpair-major} when we identify $t \cos \theta/2 \to t$ and $\Delta \sin \theta/2 \to \Delta$. The only new parameter in our Hamiltonian is the spin modulation angle $\theta$. A detailed analysis of the phase diagram of a similar model, but without magnetic field, was done recently elsewhere by the authors \cite{kesharpu-2024-propos-realiz}. We see that, even if the Hamiltonian contains the external magnetic field, it plays no important role for the \emph{x-y} spiral spin structure. Therefore, the phase diagram discussed in Ref. \cite{kesharpu-2024-propos-realiz} and the phase diagram of Eq. (\ref{eq:effectiv-ham-spiral-spin}) will be same.
The topological phase boundary can be represented as:
\begin{equation}
  \label{eq:topo-phase-spiral}
  |\mu/t|<2 \cos \left( \theta/2 \right).
\end{equation}
It means that with increase in $\theta$ the area under topological phase decreases. Physically, it can be explained by the overlap of the spin wave functions on the neighbouring sites. When The spins are along the same direction ($\theta=0$) then the overlap of wave function is highest, therefore the hopping probability is highest, $t\cos \left( \theta/2 \right) \to t$. However, when the spins are aligned along opposite directions then the overlap of the wave functions is zero, therefore the hopping probability vanishes, $t\cos \left( \theta/2 \right) \to 0$. In Eq. (\ref{eq:topo-phase-spiral}) one should keep in mind that the topologically non-trivial state is absent for $\theta=0$ (as $\Delta\sin \theta/2=0$, superconductivity is quenched) and $\theta=\pi$ (as $t\cos \theta/2$=0, kinetic energy is quenched).

\section{Application to the 2D BCS-Hubbard Model}
\begin{figure*}
  \centering
  \includegraphics[width=0.9\textwidth]{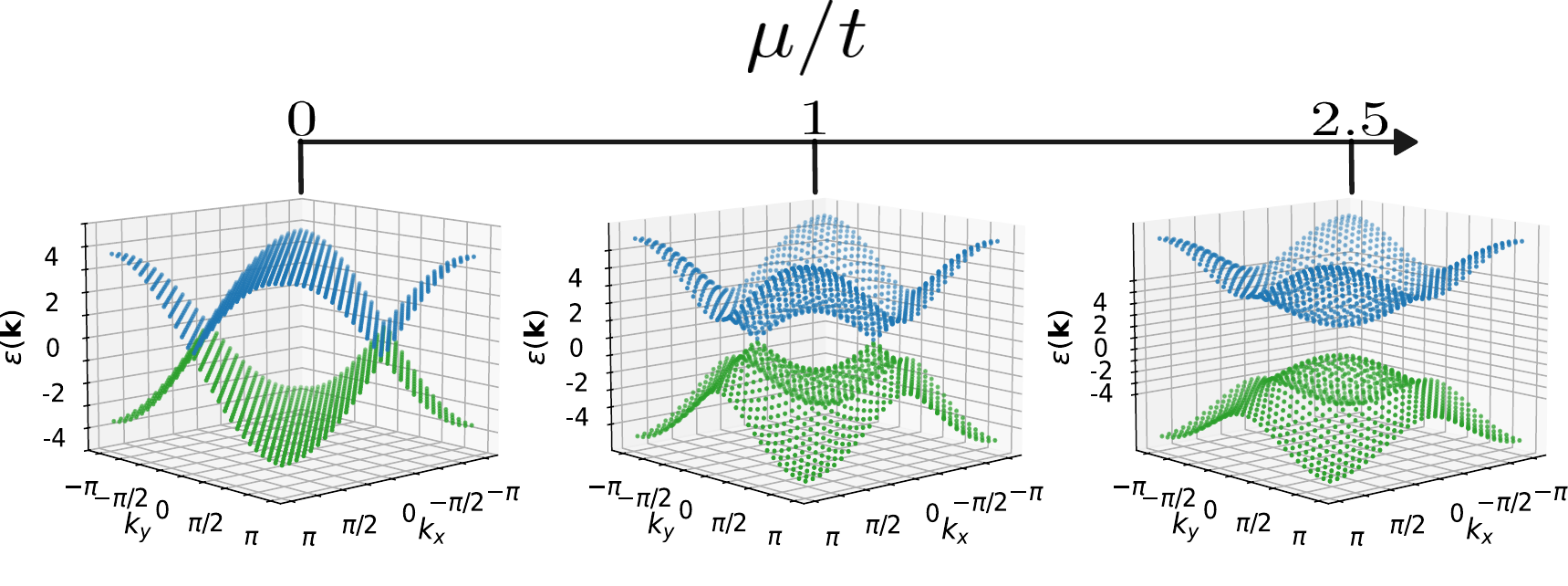}
  \caption{Plot of the energy dispersion $\epsilon_{k}$ from Eq. (\ref{eq:diag-ham-bdg}) for different values of chemical potential $\mu/t$ with constant parameters $\Delta/t=0.5$, and $\theta=\pi/4$. (left) When $\mu/t=0$ the upper (blue) and lower (green) bands touch each other along the line $k_{x}=\pm \pi + k_{y}$. (center) When $\mu/t=1$ the two bands touch each other at the points $k_{x}=\mp \arccos \left[ 1/2 \cos \left( \pi/8  \right) \right]$, $k_{y}=\pm \arccos \left[ 1/2 \cos \left( \pi/8  \right) \right]$. (right) The nodal points disappears for $\mu/t> 2 \cos \theta/2$, therefore a gap appears for $\mu/t=2.5$.}
  \label{fig:energy-disp-dep-on-mu}
\end{figure*}
Next we show the application of the method to the 2D BCS-Hubbard model under strong electronic correlation, i.e. when the on-site Coulomb repulsion $U$ is the largest energy scale ($U\to\infty$). The non-interacting BCS-Hubbard Hamiltonian is:
\begin{equation}
  \label{eq:ham-bcs-hubbard}
  \begin{aligned}
    H_{\text{BCS}}=&-t \sum\limits_{ \left\langle i,j \right\rangle, \sigma} c_{i, \sigma}^{\dagger}c_{j, \sigma} - \sum\limits_{\left\langle i,j \right\rangle,\sigma} \Delta_{ij}^{\sigma \sigma} c_{i,\sigma} c_{j,\sigma}  \\
      &\quad - \sum\limits_{\left\langle i,j \right\rangle, \sigma} \Delta_{ij}^{\sigma \sigma'} c_{i,\sigma} c_{j,\sigma'}+ \text{H.c.} + \mu \sum\limits_{ i, \sigma} c_{i, \sigma}^{\dagger}c_{i, \sigma}\\
    &\qquad \quad + U \sum\limits_{ i} \left(n_{i, \sigma} - \frac{1}{2} \right) \left(n_{i, \sigma'} - \frac{1}{2} \right).
  \end{aligned}
\end{equation}
Due to Pauli exclusion principle the $p$-wave superconducting order parameters obey $\Delta_{ij}^{\sigma \sigma}=-\Delta_{ji}^{\sigma \sigma}$, $\Delta_{ij}^{\sigma \sigma'}=\Delta_{ij}^{\sigma' \sigma}=-\Delta_{ji}^{\sigma \sigma'}$. The $\Delta_{ij}^{\sigma \sigma}$ correspond to the $m_{z}=\pm 1$ channels (triplet) and $\Delta_{ij}^{\sigma \sigma'}$ correspond to the $m_{z}=0$ channel (singlet) \cite{mineev1999-IntroductionUnconventional}. In the presence of the strong correlation ($U \to \infty$) using the mapping from Eq. (\ref{eq:Hubbard-oprs}), we can write Eq. (\ref{eq:ham-bcs-hubbard}) as:
\begin{equation}
  \label{eq:ham-bcs-hubbard-xopr}
  \begin{aligned}
    H_{\text{BCS}}=&-t \sum\limits_{\left\langle i,j \right\rangle, \sigma} X_{i}^{\sigma 0}X_{j}^{0 \sigma} - \sum\limits_{\left\langle i,j \right\rangle,\sigma} \Delta_{ij}^{\sigma \sigma} X_{i}^{0\sigma} X_{j}^{0\sigma}\\
    & -\sum\limits_{\left\langle i,j \right\rangle,\sigma} \Delta_{ij}^{\sigma \sigma'} X_{i}^{0\sigma} X_{j}^{0\sigma'}  + \text{H.c.} + \mu\sum\limits_{i, \sigma} X_{i}^{00}.
  \end{aligned}
\end{equation}
We define $\Delta_{ij}^{\uparrow \uparrow} \equiv \Delta^{+}$, $\Delta_{ij}^{\downarrow \downarrow} \equiv \Delta^{-}$, $\Delta_{ij}^{\uparrow \downarrow} = \Delta^{\downarrow \uparrow}  \equiv \Delta$, and assume $\Delta^{+}=\Delta^{-} \equiv \Delta^{\pm}$. A small note is in order regarding the defined superconducting order parameters $\Delta^{\pm}$. Physically the $\Delta^{\pm}$ represents the spin singlet pairing on the neighbouring sites, however, in the presence of the spiral spin structure one can expect $\Delta^{\pm} \approx 0$. Therefore setting $\Delta^{\pm}=0$ in Eq. (\ref{eq:ham-bcs-hubbard-xopr}), and using Eqs. (\ref{eq:fermionic-su-2-1}) -- (\ref{eq:su-2-generator-cs-symb}), the Hamiltonian in terms of the $su(2|1)$ path integral turns into:
\begin{equation}
  \label{eq:ham-bcs-hubbard-su-2}
  \begin{aligned}
    H_{\text{BCS}}&=-t \sum\limits_{\left\langle i,j \right\rangle} \bar{\xi}_{i}\xi_{j} \: a_{ij} - \Delta \sum\limits_{\left\langle i,j \right\rangle} \bar{\xi}_{i} \bar{\xi}_{j} \: \Delta_{ij}
     +\text{H.c.} + \mu \sum\limits_{i} \xi_{i} \tilde{\xi}_{i},
  \end{aligned}
\end{equation}
where,
\begin{equation}
  \label{eq:2d-bcs-hub-mod-aij}
  \begin{aligned}
    &a_{ij} \equiv \frac{1 + \bar{z}_{j} z_{i}}{\sqrt{\left( 1 + \left| z_{j} \right|^{2}\right)\left( 1 + \left| z_{i} \right|^{2}\right)}},
    \Delta_{ij} \equiv \frac{z_{j} + z_{i}}{\sqrt{\left( 1 + \left| z_{j} \right|^{2}\right)\left( 1 + \left| z_{i} \right|^{2}\right)}}.
  \end{aligned}
\end{equation}
For spiral spin field, Eq. (\ref{eq:spin-config-spiral-xy}), we have $z_{i}=\mathrm{e}^{\imath \theta_{i}}$; using this in Eq. (\ref{eq:2d-bcs-hub-mod-aij}) and substituting these results in Eq. (\ref{eq:ham-bcs-hubbard-su-2}) we will get the Hamiltonian:
\begin{equation}
  \label{eq:ham-2d-bcs-hub-mod-spiral-S-0}
  \begin{aligned}
    H_{\text{BCS}}=& - t \cos \frac{\theta}{2} \sum\limits_{\left\langle i,j \right\rangle} \bar{\xi}_{i} \xi_{j} -\Delta \cos \frac{\theta}{2} \sum\limits_{\left\langle i,j \right\rangle} \bar{\xi}_{i} \bar{\xi}_{j} + \text{H.c.}\\
    &\qquad \qquad + \mu \sum\limits_{i} \xi_{i} \bar{\xi}_{i}.
  \end{aligned}
\end{equation}
The Hamiltonian transforms to the an effective 2D Hamiltonian with $p_{x}+p_{y}$ superconducting order parameters (due to real order parameters).

To infer the topological properties one should represent the Hamiltonian in momentum space. The corresponding Hamiltonian reads:
\begin{equation}
  \label{eq:ham-k-space-bcs}
  \begin{aligned}
    H_{\text{BCS}}(\mathbf{k})
    =& -\sum\limits_{\mathbf{k}\in BZ}\bar{\xi}_{\mathbf{k}} \left[ t \cos \frac{\theta}{2} \left(\cos k_{x} + \cos k_{y} \right) \tau_{0}\right] \xi_{\mathbf{k}}\\
    & -\imath\sum\limits_{\mathbf{k}\in BZ}\ \bar{\xi}_{\mathbf{k}} \left[ \Delta \cos \frac{\theta}{2} \left(\sin k_{x} + \sin k_{y} \right) \tau_{0}\right] \bar{\xi}_{\mathbf{-k}}\\
    &+ \mu \sum\limits_{\mathbf{k}\in BZ}\bar{\xi}_{\mathbf{k}} \xi_{\mathbf{k}}.
  \end{aligned}
\end{equation}
Here $\mathbf{k}=\left( k_{x},k_{y} \right)$ represents the ordered pair of the momentum; $\xi_{\mathbf{k}}$ ($\bar{\xi}_{\mathbf{k}}$) is the electron annihilation (creation) operator for momentum $\mathbf{k}$; $\tau_{0}$ represents the zeroth Pauli matrix. The Hamiltonian can be diagonalized using Bogolyubov operators, which reads:
\begin{equation}
  \label{eq:bogolyuobv-operators}
  \gamma_{\mathbf{k}}=u_{\mathbf{k}}\xi_{\mathbf{k}}+v_{\mathbf{k}}\bar{\xi}_{-k}.
\end{equation}
$\gamma_{\mathbf{k}}$ satisfy the usual fermion anti-commutation relations; $u_{\mathbf{k}}$ and $v_{\mathbf{k}}$ are the complex coefficients. In terms of the Bogolyubov operators the diagonalized Hamiltonian reads:
\begin{widetext}
\begin{equation}
  \label{eq:diag-ham-bdg}
  \begin{aligned}
    &H_{\text{BdG}} = \sum\limits_{\mathbf{k} \in BZ} \epsilon_{\mathbf{k}} \left( \gamma_{\mathbf{k}} \bar{\gamma}_{\mathbf{k}} - \frac{1}{2}\right),
    \text{ where, }
    \epsilon_{\mathbf{k}} = \pm 2 \sqrt{\left[ t\cos \frac{\theta}{2} \left( \cos k_{x}+\cos k_{y} \right) - \mu \right]^{2} + \left[\Delta \cos \frac{\theta}{2}  \left( \sin k_{x} + \sin k_{y} \right) \right]^{2}}
  \end{aligned}
\end{equation}
\end{widetext}
The energy dispersion $\epsilon_{\mathbf{k}}$ is a function of parameters $\theta$, $\mu$ and $\Delta$. The band crossing points (nodes) are found by the zeros of $\epsilon_{k}$, in fact several possibilities are available; by tuning the parameters one can have nodal points, nodal lines, or nodal circles.

The zeros of $\epsilon_{k}$ (also known as nodes, band touching points) are found when following conditions are satisfied:
\begin{equation}
  \label{eq:condn-nodes}
  \begin{aligned}
    \cos \frac{\theta}{2} \left( \cos k_{x} + \cos k_{y}\right) = \mu/t,\\
     \left( \Delta/t \right) \:\cos \frac{\theta}{2} \left( \sin k_{x} + \sin k_{y}\right) = 0.\\
  \end{aligned}
\end{equation}
Assuming $\Delta>0$, and $ \cos \left( \theta/2 \right) \in \left( 0,1 \right)$, three different cases arise: (i) $\mu/t=0$, (ii) $  0 < \mu/t < 2\cos \left(\theta/2\right)$, (iii) $\mu/t > 2\cos \left(\theta/2\right)$. For the first case, the conditions in Eq. (\ref{eq:condn-nodes}) are satisfied for $k_{x}=\pm \pi + k_{y}$; it represents two lines on the $k_{x}$-$k_{y}$ plane. The energy dispersion for the same is shown in Fig. \ref{fig:energy-disp-dep-on-mu} (left). For the second case, i.e., $0< \mu/t <2\cos \left(\theta/2\right)$, the node points occur at
\begin{equation*}
  \label{eq:kx-ky-mu-less}
  \begin{aligned}
    &k_{x}= \mp \arccos \left[ \mu/2 t \cos \left(\theta/2\right) \right],\\
    &k_{y}= \pm \arccos \left[ \mu/2 t \cos \left(\theta/2\right) \right].
  \end{aligned}
\end{equation*}
The energy dispersion for $\mu/t=1$ is shown in Fig. \ref{fig:energy-disp-dep-on-mu} (center). One can observe two nodal points. For the third case $\mu/t>2 \cos \left( \theta/2 \right)$ no nodal points occur as shown in Fig. \ref{fig:energy-disp-dep-on-mu} (right). When the chemical potential is negative the behaviour is analogous to their positive counterparts. In other words, nodal points survive for $-2\cos \left(\theta/2\right)< \mu/t < 0$, and when $\mu/t<-2\cos \left(\theta/2\right)$ nodal points disappear (gap appears). It should be kept in mind that the energy dispersion for negative $\mu$ and positive $\mu$ are not same, as the first part of the $\epsilon_{\mathbf{k}}$ in Eq. (\ref{eq:diag-ham-bdg}) is not symmetric around $\mu=0$.

To understand the topological properties, one should first investigate the symmetries of the Hamiltonian. The time reversal symmetry ($\mathcal{T}=\mathcal{K}$) is preserved for the system; Eq. (\ref{eq:ham-k-space-bcs}) satisfies the condition $\mathcal{T}H_{\text{BCS}}(k)\mathcal{T}^{-1}=H_{\text{BCS}}(k)$. Particle-hole symmetry ($\Xi \equiv \sigma_{x}\mathcal{K}$) is also preserved for the Hamiltonian; it can be easily checked that $\Xi H_{\text{BdG}}(k) \Xi^{-1}=-H_{\text{BdG}}(k)$. Finally, the chiral symmetry ($\Pi \equiv \sigma_{x}$) is also conserved for the Hamiltonian. Hence the Hamiltonian has all three, i.e., time reversal, particle-hole, and chiral symmetries. Therefore it corresponds to the BDI class of nodal superconductors \cite{schnyder-2015-topol-surfac}.

\begin{figure}
  \centering
  \includegraphics[width=0.45\textwidth]{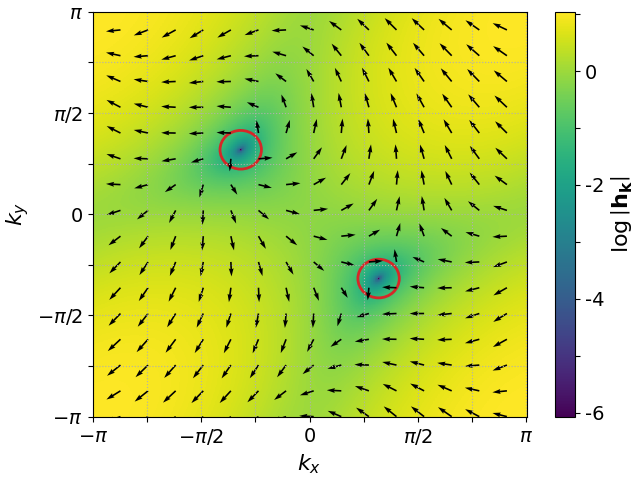}
  \caption{Plot of the components of $\mathbf{h}_{\mathbf{k}}$ from Eq. (\ref{eq:h_k-components}) for $\Delta/t=0.5$, $\mu/t=1$, and $\theta=\pi/4$. The direction of the arrows represent the relative orientation of the vector field $\mathbf{h}_{\mathbf{k}}$; it is calculated as $\atan \left( h_{\mathbf{k},y}/h_{\mathbf{k},z} \right)$. One should note $h_{\mathbf{k},z}$ represents $x$-axis and $h_{\mathbf{k},y}$ represents $y$-axis. The background color shows the logarithmic of intensity of the vector field, i.e. $\log \left| \mathbf{h}_{\mathbf{k}} \right| = \log \sqrt{ h_{\mathbf{k},x}^{2} + h_{\mathbf{k},y}^{2} + h_{\mathbf{k},z}^{2} }$. The logarithm is taken to point out the source and drain of the vector fields which occur at $k_{x}=\mp \arccos \left[ 1/2 \cos \left( \pi/8  \right) \right]$, $k_{y}=\pm \arccos \left[ 1/2 \cos \left( \pi/8  \right) \right]$. The circles around the source and drain shows the integration path for calculation of the winding number $w$ in Eq. (\ref{eq:winding-numb}).}
  \label{fig:single-hk}
\end{figure}

\begin{figure*}[tbh]
  \centering
  \includegraphics[width=0.9\textwidth]{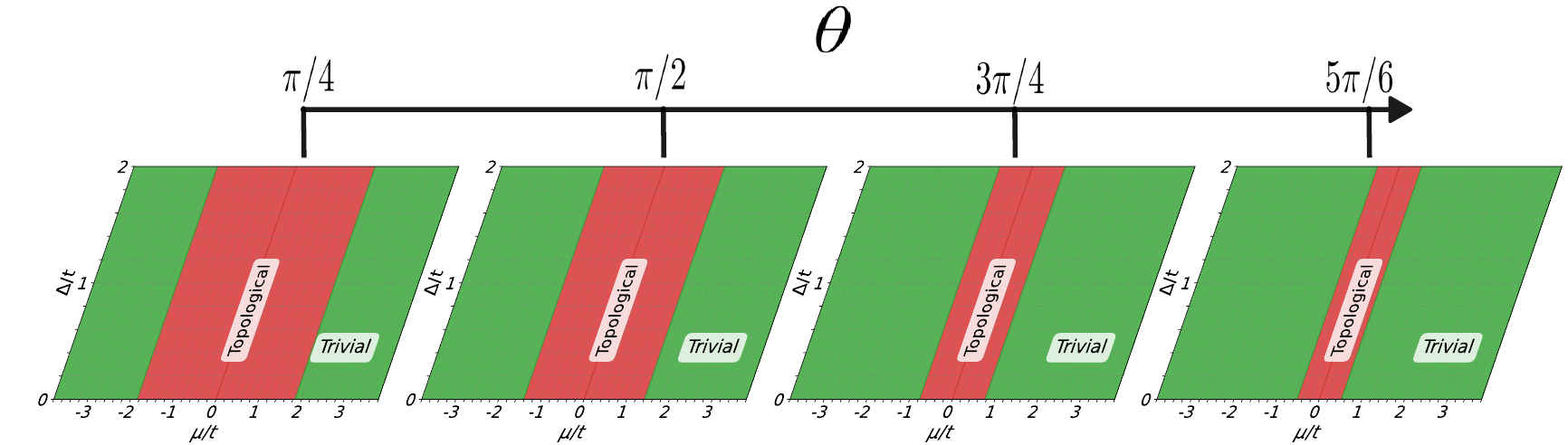}
  \caption{Topological phase diagram for BCS-Hubbard model with $p_{x}$-$p_{y}$ superconducting order parameter. The $x$-axis shows the chemical potential and the $y$-axis shows the superconducting order parameters. With increase in $\theta$ the area under topological non-trivial phase decreases. The topological condition is given in Eq. (\ref{eq:top-weyl-cond}).}
  \label{fig:top-phase-bcs}
\end{figure*}

The topological invariant for 2D nodal superconductors is the winding number. To find the winding number, first the Hamiltonian, Eq. (\ref{eq:ham-k-space-bcs}), is re written as:
\begin{equation}
  \label{eq:weyl-ham}
  \begin{aligned}
   & \begin{bmatrix}
    &\xi_{\mathbf{k}} &\bar{\xi}_{\mathbf{-k}}
    \end{bmatrix}\:
    \begin{bmatrix}
      \mathbf{h}_{\mathbf{k}} \cdot \mathbf{\tau}      
    \end{bmatrix}
    \begin{bmatrix}
      &\bar{\xi}_{\mathbf{k}} \\ &\xi_{\mathbf{-k}}
    \end{bmatrix},
  \end{aligned}
\end{equation}
where $\mathbf{h}_{\mathbf{k}}=\left( h_{\mathbf{k},x},h_{\mathbf{k},y},h_{\mathbf{k},z}\right)$ is an effective vector with components
\begin{equation}
  \label{eq:h_k-components}
  \begin{aligned}
    &h_{\mathbf{k},x}=0,\\
    &h_{\mathbf{k},y}=\Delta\cos \frac{\theta}{2} \left( \sin k_{x} + \sine k_{y} \right),\\
    &h_{\mathbf{k},z}=t\cos \frac{\theta}{2} \left( \cos k_{x} + \cos k_{y} \right) - \mu.\\
  \end{aligned}
\end{equation}
Similarly, $\mathbf{\tau}= \left( \tau_{1}, \tau_{2}, \tau_{3} \right)$ is also an effective vector, with Pauli matrices as its components. One can think of $\mathbf{h}_{\mathbf{k}}$ as a vector field on $k_{x}$-$k_{y}$ plane. The nodal points on this momentum plane are the sink or sources of the vector field $\mathbf{h}_{\mathbf{k}}$. For example in Fig. \ref{fig:single-hk} we plot the $\mathbf{h}_{\mathbf{k}}$ for parameters $\mu/t=1$, $\Delta/t=0.5$, and $\theta=\pi/4$; the sink (source) is shown by incoming (outgoing) arrows. The winding number is calculated by integrating the curl of $\mathbf{h}_{\mathbf{k}}/\left| \mathbf{h}_{\mathbf{k}} \right|$ and $\nabla_{\mathbf{k}} \mathbf{h}_{\mathbf{k}}/\left| \mathbf{h}_{\mathbf{k}} \right|$ along a closed loop around the sources and sinks. Explicitly, the winding number is found as
\begin{equation}
  \label{eq:winding-numb}
  w = \frac{1}{2 \pi \left| \mathbf{h}_{\mathbf{k}} \right|^{2}} \int\limits_{C} \left( h_{\mathbf{k},y} \: \nabla_{\mathbf{k}} h_{\mathbf{k},z} - h_{\mathbf{k},z} \: \nabla_{\mathbf{k}} h_{\mathbf{k},y} \right) dk_{x}dk_{y}.
\end{equation}
Here $\nabla_{\mathbf{k}}=\left( \partial/\partial k_{x}, \partial/\partial k_{y}, \partial/\partial k_{z} \right)$ is the gradient operator. The integration is taken along the closed contour $C$ in counterclockwise direction. In Fig. \ref{fig:single-hk} the winding number $w$ is calculated around two nodes, which is opposite to each other. These nodes are also known as the Weyl nodes. The Weyl nodes disappears only when two of them with opposite winding numbers come together. In this sense, the Weyl nodes are topologically protected. The condition for the presence of the Weyl nodes are:
\begin{equation}
  \label{eq:top-weyl-cond}
  \begin{aligned}
    &\left( \Delta/t \right) \cos \left(\frac{\theta}{2} \right) > 0, \quad \left| \mu/t \right| < 2 \cos \left(\frac{\theta}{2}  \right).
  \end{aligned}
\end{equation}
The condition Eq. (\ref{eq:top-weyl-cond}) also implicitly represent the topologically stable phase. Basically, as long as Eq. (\ref{eq:top-weyl-cond}) is satisfied for BCS Hamiltonian with $p$-wave superconducting order parameter on a square lattice. In Fig. \ref{fig:top-phase-bcs} we plot the topological phase diagram. One can observe that with increase in $\theta$ the area with non-trivial topological phase decreases.

\section{Conclusion}
\label{sec:conclusion}

Although in the previous sections the theoretical method and its application to the 1D and 2D systems have been provided, the question still remains about the applicability of the model in real materials. One experimental confirmation comes from the strongly correlated 1D NW \cite{heedt-2017-signat-inter}, where reentrant conductance behaviour was observed (when chemical potential changed monotonically). The reentrant behaviour is due to the emergence of the helical subbands induced by the Rashba spin orbit coupling and strong e-e interaction \cite{heedt-2017-signat-inter,sato-2019-stron-elect}. Analogous behaviour was also observed when the system was investigated using the method discussed in this article \cite{kesharpu-2025-reent-topol}. This fact corroborates the method at least for strongly correlated 1D cases. For the 2D cases the method can be applied to the Moire materials, as e-e correlation is strong in them \cite{balents-2020-super-stron}; it will be discussed in another work.

Regarding the perspective and usefulness of this work, the method provided here is quite general and can be applied to different Hamiltonians, given the following two conditions are satisfied: (i) e-e correlation is strong enough so that doubly occupied state can be neglected, (ii) the system should have some predefined spin structure. One can also alternatively use different numerical methods with random spin structures to investigate the properties of the systems within a spin liquid framework.

In this work we described an appropriate path integral method to treat the strongly correlated electronic systems using coherent state representation of the Hubbard operators associated with $su(2|1)$ superalgebra. The main advantage of such representation is that, the local constraint imposed by the strong electronic correlation --- either singly occupied or empty state is allowed at each atomic site --- is rigorously taken into account, unlike other approaches where this constraint is applied at the mean-field level. We showed how the $su(2|1)$ path integral method can be used to explicitly solve the Hamiltonians of these systems. One of the main advantages of the present method is the fact that the final resulting Hamiltonian is represented through fractionalized charge and spin degrees of freedom (quasi particles) of the interacting electrons. They are the so called the spinons and holons of the strongly interacting systems. As an example of that scenario we solve the strongly correlated Hamiltonians regime of the 1D Kitaev chain with extended $s$-wave superconductivity and the 2D BCS-Hubbard model on a square lattice with $p$-wave superconductivity. We showed that in these Hamiltonians the external magnetic field is not a necessary tool for the emergence of the topological superconductivity. Here the strong correlation takes the role of the external magnetic field.

\section{Acknowledgment}
\label{sec:acknowledgment}
K.K.K acknowledges the financial support from the JINR grant for young scientists and specialists, the Foundation for the Advancement of Theoretical Physics and Mathematics ”Basis” for grant \# 23-1-4-63-1. One of us, A.F., acknowledges financial support from the MEC, CNPq (Brazil) and from the Simons Foundation (USA).


\begin{thebibliography}{40}%
\makeatletter
\providecommand \@ifxundefined [1]{%
 \@ifx{#1\undefined}
}%
\providecommand \@ifnum [1]{%
 \ifnum #1\expandafter \@firstoftwo
 \else \expandafter \@secondoftwo
 \fi
}%
\providecommand \@ifx [1]{%
 \ifx #1\expandafter \@firstoftwo
 \else \expandafter \@secondoftwo
 \fi
}%
\providecommand \natexlab [1]{#1}%
\providecommand \enquote  [1]{``#1''}%
\providecommand \bibnamefont  [1]{#1}%
\providecommand \bibfnamefont [1]{#1}%
\providecommand \citenamefont [1]{#1}%
\providecommand \href@noop [0]{\@secondoftwo}%
\providecommand \href [0]{\begingroup \@sanitize@url \@href}%
\providecommand \@href[1]{\@@startlink{#1}\@@href}%
\providecommand \@@href[1]{\endgroup#1\@@endlink}%
\providecommand \@sanitize@url [0]{\catcode `\\12\catcode `\$12\catcode
  `\&12\catcode `\#12\catcode `\^12\catcode `\_12\catcode `\%12\relax}%
\providecommand \@@startlink[1]{}%
\providecommand \@@endlink[0]{}%
\providecommand \url  [0]{\begingroup\@sanitize@url \@url }%
\providecommand \@url [1]{\endgroup\@href {#1}{\urlprefix }}%
\providecommand \urlprefix  [0]{URL }%
\providecommand \Eprint [0]{\href }%
\providecommand \doibase [0]{https://doi.org/}%
\providecommand \selectlanguage [0]{\@gobble}%
\providecommand \bibinfo  [0]{\@secondoftwo}%
\providecommand \bibfield  [0]{\@secondoftwo}%
\providecommand \translation [1]{[#1]}%
\providecommand \BibitemOpen [0]{}%
\providecommand \bibitemStop [0]{}%
\providecommand \bibitemNoStop [0]{.\EOS\space}%
\providecommand \EOS [0]{\spacefactor3000\relax}%
\providecommand \BibitemShut  [1]{\csname bibitem#1\endcsname}%
\let\auto@bib@innerbib\@empty
\bibitem [{\citenamefont {Sarma}\ \emph {et~al.}(2015)\citenamefont {Sarma},
  \citenamefont {Freedman},\ and\ \citenamefont
  {Nayak}}]{sarma-2015-major-zero}%
  \BibitemOpen
  \bibfield  {author} {\bibinfo {author} {\bibfnamefont {S.~D.}\ \bibnamefont
  {Sarma}}, \bibinfo {author} {\bibfnamefont {M.}~\bibnamefont {Freedman}},\
  and\ \bibinfo {author} {\bibfnamefont {C.}~\bibnamefont {Nayak}},\ }\bibfield
   {title} {\bibinfo {title} {Majorana zero modes and topological quantum
  computation},\ }\href {https://doi.org/10.1038/npjqi.2015.1} {\bibfield
  {journal} {\bibinfo  {journal} {npj Quantum Information}\ }\textbf {\bibinfo
  {volume} {1}},\ \bibinfo {pages} {15001} (\bibinfo {year}
  {2015})}\BibitemShut {NoStop}%
\bibitem [{\citenamefont {Flensberg}\ \emph {et~al.}(2021)\citenamefont
  {Flensberg}, \citenamefont {von Oppen},\ and\ \citenamefont
  {Stern}}]{flensberg-2021-engin-platf}%
  \BibitemOpen
  \bibfield  {author} {\bibinfo {author} {\bibfnamefont {K.}~\bibnamefont
  {Flensberg}}, \bibinfo {author} {\bibfnamefont {F.}~\bibnamefont {von
  Oppen}},\ and\ \bibinfo {author} {\bibfnamefont {A.}~\bibnamefont {Stern}},\
  }\bibfield  {title} {\bibinfo {title} {Engineered platforms for topological
  superconductivity and majorana zero modes},\ }\href
  {https://doi.org/10.1038/s41578-021-00336-6} {\bibfield  {journal} {\bibinfo
  {journal} {Nature Reviews Materials}\ }\textbf {\bibinfo {volume} {6}},\
  \bibinfo {pages} {944} (\bibinfo {year} {2021})}\BibitemShut {NoStop}%
\bibitem [{\citenamefont {Frolov}\ \emph {et~al.}(2020)\citenamefont {Frolov},
  \citenamefont {Manfra},\ and\ \citenamefont {Sau}}]{frolov-2020-topol-super}%
  \BibitemOpen
  \bibfield  {author} {\bibinfo {author} {\bibfnamefont {S.~M.}\ \bibnamefont
  {Frolov}}, \bibinfo {author} {\bibfnamefont {M.~J.}\ \bibnamefont {Manfra}},\
  and\ \bibinfo {author} {\bibfnamefont {J.~D.}\ \bibnamefont {Sau}},\
  }\bibfield  {title} {\bibinfo {title} {Topological superconductivity in
  hybrid devices},\ }\href {https://doi.org/10.1038/s41567-020-0925-6}
  {\bibfield  {journal} {\bibinfo  {journal} {Nature Physics}\ }\textbf
  {\bibinfo {volume} {16}},\ \bibinfo {pages} {718} (\bibinfo {year}
  {2020})}\BibitemShut {NoStop}%
\bibitem [{\citenamefont {Avella}\ and\ \citenamefont
  {Mancini}(2012)}]{avella2012-StronglyCorrelated}%
  \BibitemOpen
  \bibinfo {editor} {\bibfnamefont {A.}~\bibnamefont {Avella}}\ and\ \bibinfo
  {editor} {\bibfnamefont {F.}~\bibnamefont {Mancini}},\ eds.,\ \href
  {https://doi.org/10.1007/978-3-642-21831-6} {\emph {\bibinfo {title}
  {Strongly {{Correlated Systems}}: {{Theoretical Methods}}}}},\ \bibinfo
  {series} {Springer {{Series}} in {{Solid-State Sciences}}}, Vol.\ \bibinfo
  {volume} {171}\ (\bibinfo  {publisher} {Springer Berlin Heidelberg},\
  \bibinfo {address} {Berlin, Heidelberg},\ \bibinfo {year} {2012})\BibitemShut
  {NoStop}%
\bibitem [{\citenamefont {S{\'e}n{\'e}chal}\ \emph {et~al.}(2004)\citenamefont
  {S{\'e}n{\'e}chal}, \citenamefont {Tremblay},\ and\ \citenamefont
  {Bourbonnais}}]{senechal2004-TheoreticalMethods}%
  \BibitemOpen
  \bibinfo {editor} {\bibfnamefont {D.}~\bibnamefont {S{\'e}n{\'e}chal}},
  \bibinfo {editor} {\bibfnamefont {A.-M.}\ \bibnamefont {Tremblay}},\ and\
  \bibinfo {editor} {\bibfnamefont {C.}~\bibnamefont {Bourbonnais}},\ eds.,\
  \href {https://doi.org/10.1007/b97552} {\emph {\bibinfo {title} {Theoretical
  {{Methods}} for {{Strongly Correlated Electrons}}}}},\ {{CRM Series}} in
  {{Mathematical Physics}}\ (\bibinfo  {publisher} {Springer-Verlag},\ \bibinfo
  {address} {New York},\ \bibinfo {year} {2004})\BibitemShut {NoStop}%
\bibitem [{\citenamefont {Tsvelik}(2001)}]{tsvelik2001-NewTheoretical}%
  \BibitemOpen
  \bibinfo {editor} {\bibfnamefont {A.~M.}\ \bibnamefont {Tsvelik}},\ ed.,\
  \href {https://doi.org/10.1007/978-94-010-0838-9} {\emph {\bibinfo {title}
  {New {{Theoretical Approaches}} to {{Strongly Correlated Systems}}}}}\
  (\bibinfo  {publisher} {Springer Netherlands},\ \bibinfo {address}
  {Dordrecht},\ \bibinfo {year} {2001})\BibitemShut {NoStop}%
\bibitem [{\citenamefont {Maciejko}\ and\ \citenamefont
  {Fiete}(2015)}]{maciejko-2015-fract-topol-insul}%
  \BibitemOpen
  \bibfield  {author} {\bibinfo {author} {\bibfnamefont {J.}~\bibnamefont
  {Maciejko}}\ and\ \bibinfo {author} {\bibfnamefont {G.~A.}\ \bibnamefont
  {Fiete}},\ }\bibfield  {title} {\bibinfo {title} {Fractionalized topological
  insulators},\ }\href {https://doi.org/10.1038/nphys3311} {\bibfield
  {journal} {\bibinfo  {journal} {Nature Physics}\ }\textbf {\bibinfo {volume}
  {11}},\ \bibinfo {pages} {385} (\bibinfo {year} {2015})}\BibitemShut
  {NoStop}%
\bibitem [{\citenamefont {Irkhin}\ and\ \citenamefont
  {Skryabin}(2022)}]{irkhin_2022_TopologicalStates_JSupercondNovMagn}%
  \BibitemOpen
  \bibfield  {author} {\bibinfo {author} {\bibfnamefont {V.~{\relax Yu}.}\
  \bibnamefont {Irkhin}}\ and\ \bibinfo {author} {\bibfnamefont {{\relax
  Yu}.~N.}\ \bibnamefont {Skryabin}},\ }\bibfield  {title} {\bibinfo {title}
  {Topological {{States}} in {{Strongly Correlated Systems}}},\ }\href
  {https://doi.org/10.1007/s10948-022-06251-3} {\bibfield  {journal} {\bibinfo
  {journal} {Journal of Superconductivity and Novel Magnetism}\ }\textbf
  {\bibinfo {volume} {35}},\ \bibinfo {pages} {2141} (\bibinfo {year}
  {2022})}\BibitemShut {NoStop}%
\bibitem [{\citenamefont {Rachel}(2018)}]{rachel-2018-inter-topol-insul}%
  \BibitemOpen
  \bibfield  {author} {\bibinfo {author} {\bibfnamefont {S.}~\bibnamefont
  {Rachel}},\ }\bibfield  {title} {\bibinfo {title} {Interacting topological
  insulators: a review},\ }\href {https://doi.org/10.1088/1361-6633/aad6a6}
  {\bibfield  {journal} {\bibinfo  {journal} {Reports on Progress in Physics}\
  }\textbf {\bibinfo {volume} {81}},\ \bibinfo {pages} {116501} (\bibinfo
  {year} {2018})}\BibitemShut {NoStop}%
\bibitem [{\citenamefont {Xie}\ \emph {et~al.}(2021)\citenamefont {Xie},
  \citenamefont {Pierce}, \citenamefont {Park}, \citenamefont {Parker},
  \citenamefont {Khalaf}, \citenamefont {Ledwith}, \citenamefont {Cao},
  \citenamefont {Lee}, \citenamefont {Chen}, \citenamefont {Forrester},
  \citenamefont {Watanabe}, \citenamefont {Taniguchi}, \citenamefont
  {Vishwanath}, \citenamefont {Jarillo-Herrero},\ and\ \citenamefont
  {Yacoby}}]{xie-2021-fract-chern}%
  \BibitemOpen
  \bibfield  {author} {\bibinfo {author} {\bibfnamefont {Y.}~\bibnamefont
  {Xie}}, \bibinfo {author} {\bibfnamefont {A.~T.}\ \bibnamefont {Pierce}},
  \bibinfo {author} {\bibfnamefont {J.~M.}\ \bibnamefont {Park}}, \bibinfo
  {author} {\bibfnamefont {D.~E.}\ \bibnamefont {Parker}}, \bibinfo {author}
  {\bibfnamefont {E.}~\bibnamefont {Khalaf}}, \bibinfo {author} {\bibfnamefont
  {P.}~\bibnamefont {Ledwith}}, \bibinfo {author} {\bibfnamefont
  {Y.}~\bibnamefont {Cao}}, \bibinfo {author} {\bibfnamefont {S.~H.}\
  \bibnamefont {Lee}}, \bibinfo {author} {\bibfnamefont {S.}~\bibnamefont
  {Chen}}, \bibinfo {author} {\bibfnamefont {P.~R.}\ \bibnamefont {Forrester}},
  \bibinfo {author} {\bibfnamefont {K.}~\bibnamefont {Watanabe}}, \bibinfo
  {author} {\bibfnamefont {T.}~\bibnamefont {Taniguchi}}, \bibinfo {author}
  {\bibfnamefont {A.}~\bibnamefont {Vishwanath}}, \bibinfo {author}
  {\bibfnamefont {P.}~\bibnamefont {Jarillo-Herrero}},\ and\ \bibinfo {author}
  {\bibfnamefont {A.}~\bibnamefont {Yacoby}},\ }\bibfield  {title} {\bibinfo
  {title} {Fractional chern insulators in magic-angle twisted bilayer
  graphene},\ }\href {https://doi.org/10.1038/s41586-021-04002-3} {\bibfield
  {journal} {\bibinfo  {journal} {Nature}\ }\textbf {\bibinfo {volume} {600}},\
  \bibinfo {pages} {439} (\bibinfo {year} {2021})}\BibitemShut {NoStop}%
\bibitem [{\citenamefont {Cai}\ \emph {et~al.}(2023)\citenamefont {Cai},
  \citenamefont {Anderson}, \citenamefont {Wang}, \citenamefont {Zhang},
  \citenamefont {Liu}, \citenamefont {Holtzmann}, \citenamefont {Zhang},
  \citenamefont {Fan}, \citenamefont {Taniguchi}, \citenamefont {Watanabe},
  \citenamefont {Ran}, \citenamefont {Cao}, \citenamefont {Fu}, \citenamefont
  {Xiao}, \citenamefont {Yao},\ and\ \citenamefont
  {Xu}}]{cai-2023-signat-fract}%
  \BibitemOpen
  \bibfield  {author} {\bibinfo {author} {\bibfnamefont {J.}~\bibnamefont
  {Cai}}, \bibinfo {author} {\bibfnamefont {E.}~\bibnamefont {Anderson}},
  \bibinfo {author} {\bibfnamefont {C.}~\bibnamefont {Wang}}, \bibinfo {author}
  {\bibfnamefont {X.}~\bibnamefont {Zhang}}, \bibinfo {author} {\bibfnamefont
  {X.}~\bibnamefont {Liu}}, \bibinfo {author} {\bibfnamefont {W.}~\bibnamefont
  {Holtzmann}}, \bibinfo {author} {\bibfnamefont {Y.}~\bibnamefont {Zhang}},
  \bibinfo {author} {\bibfnamefont {F.}~\bibnamefont {Fan}}, \bibinfo {author}
  {\bibfnamefont {T.}~\bibnamefont {Taniguchi}}, \bibinfo {author}
  {\bibfnamefont {K.}~\bibnamefont {Watanabe}}, \bibinfo {author}
  {\bibfnamefont {Y.}~\bibnamefont {Ran}}, \bibinfo {author} {\bibfnamefont
  {T.}~\bibnamefont {Cao}}, \bibinfo {author} {\bibfnamefont {L.}~\bibnamefont
  {Fu}}, \bibinfo {author} {\bibfnamefont {D.}~\bibnamefont {Xiao}}, \bibinfo
  {author} {\bibfnamefont {W.}~\bibnamefont {Yao}},\ and\ \bibinfo {author}
  {\bibfnamefont {X.}~\bibnamefont {Xu}},\ }\bibfield  {title} {\bibinfo
  {title} {Signatures of fractional quantum anomalous hall states in twisted
  mote2},\ }\href {https://doi.org/10.1038/s41586-023-06289-w} {\bibfield
  {journal} {\bibinfo  {journal} {Nature}\ }\textbf {\bibinfo {volume} {622}},\
  \bibinfo {pages} {63} (\bibinfo {year} {2023})}\BibitemShut {NoStop}%
\bibitem [{\citenamefont {Sagi}\ \emph {et~al.}(2017)\citenamefont {Sagi},
  \citenamefont {Haim}, \citenamefont {Berg}, \citenamefont {{von Oppen}},\
  and\ \citenamefont {Oreg}}]{sagi_2017_FractionalChiral_PhysRevB}%
  \BibitemOpen
  \bibfield  {author} {\bibinfo {author} {\bibfnamefont {E.}~\bibnamefont
  {Sagi}}, \bibinfo {author} {\bibfnamefont {A.}~\bibnamefont {Haim}}, \bibinfo
  {author} {\bibfnamefont {E.}~\bibnamefont {Berg}}, \bibinfo {author}
  {\bibfnamefont {F.}~\bibnamefont {{von Oppen}}},\ and\ \bibinfo {author}
  {\bibfnamefont {Y.}~\bibnamefont {Oreg}},\ }\bibfield  {title} {\bibinfo
  {title} {Fractional chiral superconductors},\ }\href
  {https://doi.org/10.1103/PhysRevB.96.235144} {\bibfield  {journal} {\bibinfo
  {journal} {Physical Review B}\ }\textbf {\bibinfo {volume} {96}},\ \bibinfo
  {pages} {235144} (\bibinfo {year} {2017})}\BibitemShut {NoStop}%
\bibitem [{\citenamefont {Laubscher}\ \emph {et~al.}(2019)\citenamefont
  {Laubscher}, \citenamefont {Loss},\ and\ \citenamefont
  {Klinovaja}}]{laubscher_2019_FractionalTopological_PhysRevRes}%
  \BibitemOpen
  \bibfield  {author} {\bibinfo {author} {\bibfnamefont {K.}~\bibnamefont
  {Laubscher}}, \bibinfo {author} {\bibfnamefont {D.}~\bibnamefont {Loss}},\
  and\ \bibinfo {author} {\bibfnamefont {J.}~\bibnamefont {Klinovaja}},\
  }\bibfield  {title} {\bibinfo {title} {Fractional topological
  superconductivity and parafermion corner states},\ }\href
  {https://doi.org/10.1103/PhysRevResearch.1.032017} {\bibfield  {journal}
  {\bibinfo  {journal} {Physical Review Research}\ }\textbf {\bibinfo {volume}
  {1}},\ \bibinfo {pages} {032017} (\bibinfo {year} {2019})}\BibitemShut
  {NoStop}%
\bibitem [{\citenamefont {Ronetti}\ \emph {et~al.}(2021)\citenamefont
  {Ronetti}, \citenamefont {Loss},\ and\ \citenamefont
  {Klinovaja}}]{ronetti_2021_ClockModel_PhysRevB}%
  \BibitemOpen
  \bibfield  {author} {\bibinfo {author} {\bibfnamefont {F.}~\bibnamefont
  {Ronetti}}, \bibinfo {author} {\bibfnamefont {D.}~\bibnamefont {Loss}},\ and\
  \bibinfo {author} {\bibfnamefont {J.}~\bibnamefont {Klinovaja}},\ }\bibfield
  {title} {\bibinfo {title} {Clock model and parafermions in {{Rashba}}
  nanowires},\ }\href {https://doi.org/10.1103/PhysRevB.103.235410} {\bibfield
  {journal} {\bibinfo  {journal} {Physical Review B}\ }\textbf {\bibinfo
  {volume} {103}},\ \bibinfo {pages} {235410} (\bibinfo {year}
  {2021})}\BibitemShut {NoStop}%
\bibitem [{\citenamefont {Del~Pozo}\ \emph {et~al.}(2023)\citenamefont
  {Del~Pozo}, \citenamefont {Herviou},\ and\ \citenamefont
  {Le~Hur}}]{delpozo_2023_FractionalTopology_PhysRevB}%
  \BibitemOpen
  \bibfield  {author} {\bibinfo {author} {\bibfnamefont {F.}~\bibnamefont
  {Del~Pozo}}, \bibinfo {author} {\bibfnamefont {L.}~\bibnamefont {Herviou}},\
  and\ \bibinfo {author} {\bibfnamefont {K.}~\bibnamefont {Le~Hur}},\
  }\bibfield  {title} {\bibinfo {title} {Fractional topology in interacting
  one-dimensional superconductors},\ }\href
  {https://doi.org/10.1103/PhysRevB.107.155134} {\bibfield  {journal} {\bibinfo
   {journal} {Physical Review B}\ }\textbf {\bibinfo {volume} {107}},\ \bibinfo
  {pages} {155134} (\bibinfo {year} {2023})}\BibitemShut {NoStop}%
\bibitem [{\citenamefont {Schnyder}\ and\ \citenamefont
  {Brydon}(2015)}]{schnyder-2015-topol-surfac}%
  \BibitemOpen
  \bibfield  {author} {\bibinfo {author} {\bibfnamefont {A.~P.}\ \bibnamefont
  {Schnyder}}\ and\ \bibinfo {author} {\bibfnamefont {P.~M.~R.}\ \bibnamefont
  {Brydon}},\ }\bibfield  {title} {\bibinfo {title} {Topological surface states
  in nodal superconductors},\ }\href
  {https://doi.org/10.1088/0953-8984/27/24/243201} {\bibfield  {journal}
  {\bibinfo  {journal} {Journal of Physics: Condensed Matter}\ }\textbf
  {\bibinfo {volume} {27}},\ \bibinfo {pages} {243201} (\bibinfo {year}
  {2015})}\BibitemShut {NoStop}%
\bibitem [{\citenamefont {Schnyder}\ \emph {et~al.}(2012)\citenamefont
  {Schnyder}, \citenamefont {Brydon},\ and\ \citenamefont
  {Timm}}]{schnyder-2012-types-topol}%
  \BibitemOpen
  \bibfield  {author} {\bibinfo {author} {\bibfnamefont {A.~P.}\ \bibnamefont
  {Schnyder}}, \bibinfo {author} {\bibfnamefont {P.~M.~R.}\ \bibnamefont
  {Brydon}},\ and\ \bibinfo {author} {\bibfnamefont {C.}~\bibnamefont {Timm}},\
  }\bibfield  {title} {\bibinfo {title} {Types of topological surface states in
  nodal noncentrosymmetric superconductors},\ }\href
  {https://doi.org/10.1103/physrevb.85.024522} {\bibfield  {journal} {\bibinfo
  {journal} {Physical Review B}\ }\textbf {\bibinfo {volume} {85}},\ \bibinfo
  {pages} {024522} (\bibinfo {year} {2012})}\BibitemShut {NoStop}%
\bibitem [{\citenamefont {Zhang}\ \emph {et~al.}(2019)\citenamefont {Zhang},
  \citenamefont {Cole}, \citenamefont {Wu},\ and\ \citenamefont
  {Sarma}}]{zhang-2019-higher-order}%
  \BibitemOpen
  \bibfield  {author} {\bibinfo {author} {\bibfnamefont {R.-X.}\ \bibnamefont
  {Zhang}}, \bibinfo {author} {\bibfnamefont {W.~S.}\ \bibnamefont {Cole}},
  \bibinfo {author} {\bibfnamefont {X.}~\bibnamefont {Wu}},\ and\ \bibinfo
  {author} {\bibfnamefont {S.~D.}\ \bibnamefont {Sarma}},\ }\bibfield  {title}
  {\bibinfo {title} {Higher-order topology and nodal topological
  superconductivity in fe(se,te) heterostructures},\ }\href
  {https://doi.org/10.1103/physrevlett.123.167001} {\bibfield  {journal}
  {\bibinfo  {journal} {Physical Review Letters}\ }\textbf {\bibinfo {volume}
  {123}},\ \bibinfo {pages} {167001} (\bibinfo {year} {2019})}\BibitemShut
  {NoStop}%
\bibitem [{\citenamefont {Nayak}\ \emph {et~al.}(2021)\citenamefont {Nayak},
  \citenamefont {Steinbok}, \citenamefont {Roet}, \citenamefont {Koo},
  \citenamefont {Margalit}, \citenamefont {Feldman}, \citenamefont {Almoalem},
  \citenamefont {Kanigel}, \citenamefont {Fiete}, \citenamefont {Yan},
  \citenamefont {Oreg}, \citenamefont {Avraham},\ and\ \citenamefont
  {Beidenkopf}}]{nayak-2021-eviden-topol}%
  \BibitemOpen
  \bibfield  {author} {\bibinfo {author} {\bibfnamefont {A.~K.}\ \bibnamefont
  {Nayak}}, \bibinfo {author} {\bibfnamefont {A.}~\bibnamefont {Steinbok}},
  \bibinfo {author} {\bibfnamefont {Y.}~\bibnamefont {Roet}}, \bibinfo {author}
  {\bibfnamefont {J.}~\bibnamefont {Koo}}, \bibinfo {author} {\bibfnamefont
  {G.}~\bibnamefont {Margalit}}, \bibinfo {author} {\bibfnamefont
  {I.}~\bibnamefont {Feldman}}, \bibinfo {author} {\bibfnamefont
  {A.}~\bibnamefont {Almoalem}}, \bibinfo {author} {\bibfnamefont
  {A.}~\bibnamefont {Kanigel}}, \bibinfo {author} {\bibfnamefont {G.~A.}\
  \bibnamefont {Fiete}}, \bibinfo {author} {\bibfnamefont {B.}~\bibnamefont
  {Yan}}, \bibinfo {author} {\bibfnamefont {Y.}~\bibnamefont {Oreg}}, \bibinfo
  {author} {\bibfnamefont {N.}~\bibnamefont {Avraham}},\ and\ \bibinfo {author}
  {\bibfnamefont {H.}~\bibnamefont {Beidenkopf}},\ }\bibfield  {title}
  {\bibinfo {title} {Evidence of topological boundary modes with topological
  nodal-point superconductivity},\ }\href
  {https://doi.org/10.1038/s41567-021-01376-z} {\bibfield  {journal} {\bibinfo
  {journal} {Nature Physics}\ }\textbf {\bibinfo {volume} {17}},\ \bibinfo
  {pages} {1413} (\bibinfo {year} {2021})}\BibitemShut {NoStop}%
\bibitem [{\citenamefont {He}\ \emph {et~al.}(2018)\citenamefont {He},
  \citenamefont {Zhou}, \citenamefont {He}, \citenamefont {Yuan}, \citenamefont
  {Zhang},\ and\ \citenamefont {Law}}]{he-2018-magnet-field}%
  \BibitemOpen
  \bibfield  {author} {\bibinfo {author} {\bibfnamefont {W.-Y.}\ \bibnamefont
  {He}}, \bibinfo {author} {\bibfnamefont {B.~T.}\ \bibnamefont {Zhou}},
  \bibinfo {author} {\bibfnamefont {J.~J.}\ \bibnamefont {He}}, \bibinfo
  {author} {\bibfnamefont {N.~F.~Q.}\ \bibnamefont {Yuan}}, \bibinfo {author}
  {\bibfnamefont {T.}~\bibnamefont {Zhang}},\ and\ \bibinfo {author}
  {\bibfnamefont {K.~T.}\ \bibnamefont {Law}},\ }\bibfield  {title} {\bibinfo
  {title} {Magnetic field driven nodal topological superconductivity in
  monolayer transition metal dichalcogenides},\ }\href
  {https://doi.org/10.1038/s42005-018-0041-4} {\bibfield  {journal} {\bibinfo
  {journal} {Communications Physics}\ }\textbf {\bibinfo {volume} {1}},\
  \bibinfo {pages} {40} (\bibinfo {year} {2018})}\BibitemShut {NoStop}%
\bibitem [{\citenamefont {Lu}\ \emph {et~al.}(2024)\citenamefont {Lu},
  \citenamefont {Reddy}, \citenamefont {Jeon}, \citenamefont {Mazza},
  \citenamefont {Brahlek}, \citenamefont {Wu}, \citenamefont {Yang},
  \citenamefont {Cook}, \citenamefont {Conner}, \citenamefont {Zhang},
  \citenamefont {Chakraborty}, \citenamefont {Yao}, \citenamefont {Tien},
  \citenamefont {Tseng}, \citenamefont {Yang}, \citenamefont {Lien},
  \citenamefont {Lin}, \citenamefont {Chiang}, \citenamefont {Vignale},
  \citenamefont {Li}, \citenamefont {Chang}, \citenamefont {Moore},\ and\
  \citenamefont {Bian}}]{lu-2024-realiz-two}%
  \BibitemOpen
  \bibfield  {author} {\bibinfo {author} {\bibfnamefont {Q.}~\bibnamefont
  {Lu}}, \bibinfo {author} {\bibfnamefont {P.~V.~S.}\ \bibnamefont {Reddy}},
  \bibinfo {author} {\bibfnamefont {H.}~\bibnamefont {Jeon}}, \bibinfo {author}
  {\bibfnamefont {A.~R.}\ \bibnamefont {Mazza}}, \bibinfo {author}
  {\bibfnamefont {M.}~\bibnamefont {Brahlek}}, \bibinfo {author} {\bibfnamefont
  {W.}~\bibnamefont {Wu}}, \bibinfo {author} {\bibfnamefont {S.~A.}\
  \bibnamefont {Yang}}, \bibinfo {author} {\bibfnamefont {J.}~\bibnamefont
  {Cook}}, \bibinfo {author} {\bibfnamefont {C.}~\bibnamefont {Conner}},
  \bibinfo {author} {\bibfnamefont {X.}~\bibnamefont {Zhang}}, \bibinfo
  {author} {\bibfnamefont {A.}~\bibnamefont {Chakraborty}}, \bibinfo {author}
  {\bibfnamefont {Y.-T.}\ \bibnamefont {Yao}}, \bibinfo {author} {\bibfnamefont
  {H.-J.}\ \bibnamefont {Tien}}, \bibinfo {author} {\bibfnamefont {C.-H.}\
  \bibnamefont {Tseng}}, \bibinfo {author} {\bibfnamefont {P.-Y.}\ \bibnamefont
  {Yang}}, \bibinfo {author} {\bibfnamefont {S.-W.}\ \bibnamefont {Lien}},
  \bibinfo {author} {\bibfnamefont {H.}~\bibnamefont {Lin}}, \bibinfo {author}
  {\bibfnamefont {T.-C.}\ \bibnamefont {Chiang}}, \bibinfo {author}
  {\bibfnamefont {G.}~\bibnamefont {Vignale}}, \bibinfo {author} {\bibfnamefont
  {A.-P.}\ \bibnamefont {Li}}, \bibinfo {author} {\bibfnamefont {T.-R.}\
  \bibnamefont {Chang}}, \bibinfo {author} {\bibfnamefont {R.~G.}\ \bibnamefont
  {Moore}},\ and\ \bibinfo {author} {\bibfnamefont {G.}~\bibnamefont {Bian}},\
  }\bibfield  {title} {\bibinfo {title} {Realization of a two-dimensional weyl
  semimetal and topological fermi strings},\ }\href
  {https://doi.org/10.1038/s41467-024-50329-6} {\bibfield  {journal} {\bibinfo
  {journal} {Nature Communications}\ }\textbf {\bibinfo {volume} {15}},\
  \bibinfo {pages} {6001} (\bibinfo {year} {2024})}\BibitemShut {NoStop}%
\bibitem [{\citenamefont {Kesharpu}\ \emph {et~al.}(2025)\citenamefont
  {Kesharpu}, \citenamefont {Kochetov},\ and\ \citenamefont
  {Ferraz}}]{kesharpu-2025-reent-topol}%
  \BibitemOpen
  \bibfield  {author} {\bibinfo {author} {\bibfnamefont {K.~K.}\ \bibnamefont
  {Kesharpu}}, \bibinfo {author} {\bibfnamefont {E.~A.}\ \bibnamefont
  {Kochetov}},\ and\ \bibinfo {author} {\bibfnamefont {A.}~\bibnamefont
  {Ferraz}},\ }\bibfield  {title} {\bibinfo {title} {Reentrant topological
  order in a strongly correlated nanowire due to rashba spin-orbit coupling},\
  }\href {https://doi.org/10.1103/physrevb.111.115153} {\bibfield  {journal}
  {\bibinfo  {journal} {Physical Review B}\ }\textbf {\bibinfo {volume}
  {111}},\ \bibinfo {pages} {115153} (\bibinfo {year} {2025})}\BibitemShut
  {NoStop}%
\bibitem [{\citenamefont {Kesharpu}\ \emph {et~al.}(2024)\citenamefont
  {Kesharpu}, \citenamefont {Kochetov},\ and\ \citenamefont
  {Ferraz}}]{kesharpu-2024-propos-realiz}%
  \BibitemOpen
  \bibfield  {author} {\bibinfo {author} {\bibfnamefont {K.~K.}\ \bibnamefont
  {Kesharpu}}, \bibinfo {author} {\bibfnamefont {E.~A.}\ \bibnamefont
  {Kochetov}},\ and\ \bibinfo {author} {\bibfnamefont {A.}~\bibnamefont
  {Ferraz}},\ }\bibfield  {title} {\bibinfo {title} {Proposal for realizing
  majorana fermions without external magnetic field in strongly correlated
  nanowires},\ }\href {https://doi.org/10.1103/physrevb.109.115140} {\bibfield
  {journal} {\bibinfo  {journal} {Physical Review B}\ }\textbf {\bibinfo
  {volume} {109}},\ \bibinfo {pages} {115140} (\bibinfo {year}
  {2024})}\BibitemShut {NoStop}%
\bibitem [{\citenamefont
  {Kesharpu}(2024)}]{kesharpu_2024_TopologicalHall_PhysRevB}%
  \BibitemOpen
  \bibfield  {author} {\bibinfo {author} {\bibfnamefont {K.~K.}\ \bibnamefont
  {Kesharpu}},\ }\bibfield  {title} {\bibinfo {title} {Topological {{Hall}}
  effect in strongly correlated layered magnets: {{The}} effect of the spin of
  the magnetic atoms and of the polar and azimuthal angles subtended by the
  spin texture},\ }\href {https://doi.org/10.1103/PhysRevB.109.205120}
  {\bibfield  {journal} {\bibinfo  {journal} {Physical Review B}\ }\textbf
  {\bibinfo {volume} {109}},\ \bibinfo {pages} {205120} (\bibinfo {year}
  {2024})}\BibitemShut {NoStop}%
\bibitem [{\citenamefont {Kesharpu}\ \emph {et~al.}(2023)\citenamefont
  {Kesharpu}, \citenamefont {Kochetov},\ and\ \citenamefont
  {Ferraz}}]{kesharpu-2023-topol-hall}%
  \BibitemOpen
  \bibfield  {author} {\bibinfo {author} {\bibfnamefont {K.~K.}\ \bibnamefont
  {Kesharpu}}, \bibinfo {author} {\bibfnamefont {E.~A.}\ \bibnamefont
  {Kochetov}},\ and\ \bibinfo {author} {\bibfnamefont {A.}~\bibnamefont
  {Ferraz}},\ }\bibfield  {title} {\bibinfo {title} {Topological hall effect
  induced by classical large-spin background: $su(2|1)$ path-integral
  approach},\ }\href {https://doi.org/10.1103/physrevb.107.155146} {\bibfield
  {journal} {\bibinfo  {journal} {Physical Review B}\ }\textbf {\bibinfo
  {volume} {107}},\ \bibinfo {pages} {155146} (\bibinfo {year}
  {2023})}\BibitemShut {NoStop}%
\bibitem [{\citenamefont {Ferraz}\ and\ \citenamefont
  {Kochetov}(2011)}]{ferraz-2011-effec-action}%
  \BibitemOpen
  \bibfield  {author} {\bibinfo {author} {\bibfnamefont {A.}~\bibnamefont
  {Ferraz}}\ and\ \bibinfo {author} {\bibfnamefont {E.}~\bibnamefont
  {Kochetov}},\ }\bibfield  {title} {\bibinfo {title} {Effective action for
  strongly correlated electron systems},\ }\href
  {https://doi.org/10.1016/j.nuclphysb.2011.08.011} {\bibfield  {journal}
  {\bibinfo  {journal} {Nuclear Physics B}\ }\textbf {\bibinfo {volume}
  {853}},\ \bibinfo {pages} {710} (\bibinfo {year} {2011})}\BibitemShut
  {NoStop}%
\bibitem [{\citenamefont {Ferraz}\ and\ \citenamefont
  {Kochetov}(2023)}]{ferraz-2023-connec-between}%
  \BibitemOpen
  \bibfield  {author} {\bibinfo {author} {\bibfnamefont {A.}~\bibnamefont
  {Ferraz}}\ and\ \bibinfo {author} {\bibfnamefont {E.}~\bibnamefont
  {Kochetov}},\ }\bibfield  {title} {\bibinfo {title} {Connection between the
  kitaev chain and the gutzwiller-projected bcs model},\ }\href
  {https://doi.org/10.1016/j.aop.2023.169234} {\bibfield  {journal} {\bibinfo
  {journal} {Annals of Physics}\ }\textbf {\bibinfo {volume} {456}},\ \bibinfo
  {pages} {169234} (\bibinfo {year} {2023})}\BibitemShut {NoStop}%
\bibitem [{\citenamefont {Wiegmann}(1988)}]{wiegmann-1988-super-stron}%
  \BibitemOpen
  \bibfield  {author} {\bibinfo {author} {\bibfnamefont {P.~B.}\ \bibnamefont
  {Wiegmann}},\ }\bibfield  {title} {\bibinfo {title} {Superconductivity in
  strongly correlated electronic systems and confinement versus deconfinement
  phenomenon},\ }\href {https://doi.org/10.1103/physrevlett.60.821} {\bibfield
  {journal} {\bibinfo  {journal} {Physical Review Letters}\ }\textbf {\bibinfo
  {volume} {60}},\ \bibinfo {pages} {821} (\bibinfo {year} {1988})}\BibitemShut
  {NoStop}%
\bibitem [{\citenamefont {Stone}(1989)}]{stone-1989-super-quant}%
  \BibitemOpen
  \bibfield  {author} {\bibinfo {author} {\bibfnamefont {M.}~\bibnamefont
  {Stone}},\ }\bibfield  {title} {\bibinfo {title} {Supersymmetry and the
  quantum mechanics of spin},\ }\href
  {https://doi.org/10.1016/0550-3213(89)90408-2} {\bibfield  {journal}
  {\bibinfo  {journal} {Nuclear Physics B}\ }\textbf {\bibinfo {volume}
  {314}},\ \bibinfo {pages} {557} (\bibinfo {year} {1989})}\BibitemShut
  {NoStop}%
\bibitem [{\citenamefont {Ogata}\ and\ \citenamefont
  {Shiba}(1990)}]{ogata-1990-bethe-ansat}%
  \BibitemOpen
  \bibfield  {author} {\bibinfo {author} {\bibfnamefont {M.}~\bibnamefont
  {Ogata}}\ and\ \bibinfo {author} {\bibfnamefont {H.}~\bibnamefont {Shiba}},\
  }\bibfield  {title} {\bibinfo {title} {Bethe-ansatz wave function, momentum
  distribution, and spin correlation in the one-dimensional strongly correlated
  hubbard model},\ }\href {https://doi.org/10.1103/physrevb.41.2326} {\bibfield
   {journal} {\bibinfo  {journal} {Physical Review B}\ }\textbf {\bibinfo
  {volume} {41}},\ \bibinfo {pages} {2326} (\bibinfo {year}
  {1990})}\BibitemShut {NoStop}%
\bibitem [{\citenamefont {Nakahara}(2018)}]{nakahara2018-GeometryTopology}%
  \BibitemOpen
  \bibfield  {author} {\bibinfo {author} {\bibfnamefont {M.}~\bibnamefont
  {Nakahara}},\ }\href {https://doi.org/10.1201/9781315275826} {\emph {\bibinfo
  {title} {Geometry, {{Topology}} and {{Physics}}}}},\ \bibinfo {edition}
  {2nd}\ ed.\ (\bibinfo  {publisher} {CRC Press},\ \bibinfo {year}
  {2018})\BibitemShut {NoStop}%
\bibitem [{\citenamefont {Oreg}\ \emph {et~al.}(2010)\citenamefont {Oreg},
  \citenamefont {Refael},\ and\ \citenamefont {von
  Oppen}}]{oreg-2010-helic-liquid}%
  \BibitemOpen
  \bibfield  {author} {\bibinfo {author} {\bibfnamefont {Y.}~\bibnamefont
  {Oreg}}, \bibinfo {author} {\bibfnamefont {G.}~\bibnamefont {Refael}},\ and\
  \bibinfo {author} {\bibfnamefont {F.}~\bibnamefont {von Oppen}},\ }\bibfield
  {title} {\bibinfo {title} {Helical liquids and majorana bound states in
  quantum wires},\ }\href {https://doi.org/10.1103/physrevlett.105.177002}
  {\bibfield  {journal} {\bibinfo  {journal} {Physical Review Letters}\
  }\textbf {\bibinfo {volume} {105}},\ \bibinfo {pages} {177002} (\bibinfo
  {year} {2010})}\BibitemShut {NoStop}%
\bibitem [{\citenamefont {Lutchyn}\ \emph {et~al.}(2018)\citenamefont
  {Lutchyn}, \citenamefont {Bakkers}, \citenamefont {Kouwenhoven},
  \citenamefont {Krogstrup}, \citenamefont {Marcus},\ and\ \citenamefont
  {Oreg}}]{lutchyn-2018-major-zero}%
  \BibitemOpen
  \bibfield  {author} {\bibinfo {author} {\bibfnamefont {R.~M.}\ \bibnamefont
  {Lutchyn}}, \bibinfo {author} {\bibfnamefont {E.~P. A.~M.}\ \bibnamefont
  {Bakkers}}, \bibinfo {author} {\bibfnamefont {L.~P.}\ \bibnamefont
  {Kouwenhoven}}, \bibinfo {author} {\bibfnamefont {P.}~\bibnamefont
  {Krogstrup}}, \bibinfo {author} {\bibfnamefont {C.~M.}\ \bibnamefont
  {Marcus}},\ and\ \bibinfo {author} {\bibfnamefont {Y.}~\bibnamefont {Oreg}},\
  }\bibfield  {title} {\bibinfo {title} {Majorana zero modes in
  superconductor-semiconductor heterostructures},\ }\href
  {https://doi.org/10.1038/s41578-018-0003-1} {\bibfield  {journal} {\bibinfo
  {journal} {Nature Reviews Materials}\ }\textbf {\bibinfo {volume} {3}},\
  \bibinfo {pages} {52} (\bibinfo {year} {2018})}\BibitemShut {NoStop}%
\bibitem [{\citenamefont {Streda}\ and\ \citenamefont
  {Seba}(2003)}]{streda-2003-antis-spin}%
  \BibitemOpen
  \bibfield  {author} {\bibinfo {author} {\bibfnamefont {P.}~\bibnamefont
  {Streda}}\ and\ \bibinfo {author} {\bibfnamefont {P.}~\bibnamefont {Seba}},\
  }\bibfield  {title} {\bibinfo {title} {Antisymmetric spin filtering in
  one-dimensional electron systems with uniform spin-orbit coupling},\ }\href
  {https://doi.org/10.1103/physrevlett.90.256601} {\bibfield  {journal}
  {\bibinfo  {journal} {Physical Review Letters}\ }\textbf {\bibinfo {volume}
  {90}},\ \bibinfo {pages} {256601} (\bibinfo {year} {2003})}\BibitemShut
  {NoStop}%
\bibitem [{\citenamefont {Kitaev}(2001)}]{kitaev-2001-unpair-major}%
  \BibitemOpen
  \bibfield  {author} {\bibinfo {author} {\bibfnamefont {A.~Y.}\ \bibnamefont
  {Kitaev}},\ }\bibfield  {title} {\bibinfo {title} {Unpaired majorana fermions
  in quantum wires},\ }\href {https://doi.org/10.1070/1063-7869/44/10s/s29}
  {\bibfield  {journal} {\bibinfo  {journal} {Physics-Uspekhi}\ }\textbf
  {\bibinfo {volume} {44}},\ \bibinfo {pages} {131} (\bibinfo {year}
  {2001})}\BibitemShut {NoStop}%
\bibitem [{\citenamefont {Alicea}(2012)}]{alicea-2012-new-direc}%
  \BibitemOpen
  \bibfield  {author} {\bibinfo {author} {\bibfnamefont {J.}~\bibnamefont
  {Alicea}},\ }\bibfield  {title} {\bibinfo {title} {New directions in the
  pursuit of majorana fermions in solid state systems},\ }\href
  {https://doi.org/10.1088/0034-4885/75/7/076501} {\bibfield  {journal}
  {\bibinfo  {journal} {Reports on Progress in Physics}\ }\textbf {\bibinfo
  {volume} {75}},\ \bibinfo {pages} {076501} (\bibinfo {year}
  {2012})}\BibitemShut {NoStop}%
\bibitem [{\citenamefont {Mineev}\ and\ \citenamefont
  {Samokhin}(1999)}]{mineev1999-IntroductionUnconventional}%
  \BibitemOpen
  \bibfield  {author} {\bibinfo {author} {\bibfnamefont {V.~P.}\ \bibnamefont
  {Mineev}}\ and\ \bibinfo {author} {\bibfnamefont {K.}~\bibnamefont
  {Samokhin}},\ }\href@noop {} {\emph {\bibinfo {title} {Introduction to
  {{Unconventional Superconductivity}}}}}\ (\bibinfo  {publisher} {CRC Press},\
  \bibinfo {year} {1999})\BibitemShut {NoStop}%
\bibitem [{\citenamefont {Heedt}\ \emph {et~al.}(2017)\citenamefont {Heedt},
  \citenamefont {Ziani}, \citenamefont {Cr{\'e}pin}, \citenamefont {Prost},
  \citenamefont {Trellenkamp}, \citenamefont {Schubert}, \citenamefont
  {Gr{\"u}tzmacher}, \citenamefont {Trauzettel},\ and\ \citenamefont
  {Sch{\"a}pers}}]{heedt-2017-signat-inter}%
  \BibitemOpen
  \bibfield  {author} {\bibinfo {author} {\bibfnamefont {S.}~\bibnamefont
  {Heedt}}, \bibinfo {author} {\bibfnamefont {N.~T.}\ \bibnamefont {Ziani}},
  \bibinfo {author} {\bibfnamefont {F.}~\bibnamefont {Cr{\'e}pin}}, \bibinfo
  {author} {\bibfnamefont {W.}~\bibnamefont {Prost}}, \bibinfo {author}
  {\bibfnamefont {S.}~\bibnamefont {Trellenkamp}}, \bibinfo {author}
  {\bibfnamefont {J.}~\bibnamefont {Schubert}}, \bibinfo {author}
  {\bibfnamefont {D.}~\bibnamefont {Gr{\"u}tzmacher}}, \bibinfo {author}
  {\bibfnamefont {B.}~\bibnamefont {Trauzettel}},\ and\ \bibinfo {author}
  {\bibfnamefont {T.}~\bibnamefont {Sch{\"a}pers}},\ }\bibfield  {title}
  {\bibinfo {title} {Signatures of interaction-induced helical gaps in nanowire
  quantum point contacts},\ }\href {https://doi.org/10.1038/nphys4070}
  {\bibfield  {journal} {\bibinfo  {journal} {Nature Physics}\ }\textbf
  {\bibinfo {volume} {13}},\ \bibinfo {pages} {563} (\bibinfo {year}
  {2017})}\BibitemShut {NoStop}%
\bibitem [{\citenamefont {Sato}\ \emph {et~al.}(2019)\citenamefont {Sato},
  \citenamefont {Matsuo}, \citenamefont {Hsu}, \citenamefont {Stano},
  \citenamefont {Ueda}, \citenamefont {Takeshige}, \citenamefont {Kamata},
  \citenamefont {Lee}, \citenamefont {Shojaei}, \citenamefont {Wickramasinghe},
  \citenamefont {Shabani}, \citenamefont {Palmstr{\o}m}, \citenamefont
  {Tokura}, \citenamefont {Loss},\ and\ \citenamefont
  {Tarucha}}]{sato-2019-stron-elect}%
  \BibitemOpen
  \bibfield  {author} {\bibinfo {author} {\bibfnamefont {Y.}~\bibnamefont
  {Sato}}, \bibinfo {author} {\bibfnamefont {S.}~\bibnamefont {Matsuo}},
  \bibinfo {author} {\bibfnamefont {C.-H.}\ \bibnamefont {Hsu}}, \bibinfo
  {author} {\bibfnamefont {P.}~\bibnamefont {Stano}}, \bibinfo {author}
  {\bibfnamefont {K.}~\bibnamefont {Ueda}}, \bibinfo {author} {\bibfnamefont
  {Y.}~\bibnamefont {Takeshige}}, \bibinfo {author} {\bibfnamefont
  {H.}~\bibnamefont {Kamata}}, \bibinfo {author} {\bibfnamefont {J.~S.}\
  \bibnamefont {Lee}}, \bibinfo {author} {\bibfnamefont {B.}~\bibnamefont
  {Shojaei}}, \bibinfo {author} {\bibfnamefont {K.}~\bibnamefont
  {Wickramasinghe}}, \bibinfo {author} {\bibfnamefont {J.}~\bibnamefont
  {Shabani}}, \bibinfo {author} {\bibfnamefont {C.}~\bibnamefont
  {Palmstr{\o}m}}, \bibinfo {author} {\bibfnamefont {Y.}~\bibnamefont
  {Tokura}}, \bibinfo {author} {\bibfnamefont {D.}~\bibnamefont {Loss}},\ and\
  \bibinfo {author} {\bibfnamefont {S.}~\bibnamefont {Tarucha}},\ }\bibfield
  {title} {\bibinfo {title} {Strong electron-electron interactions of a
  tomonaga-luttinger liquid observed in inas quantum wires},\ }\href
  {https://doi.org/10.1103/physrevb.99.155304} {\bibfield  {journal} {\bibinfo
  {journal} {Physical Review B}\ }\textbf {\bibinfo {volume} {99}},\ \bibinfo
  {pages} {155304} (\bibinfo {year} {2019})}\BibitemShut {NoStop}%
\bibitem [{\citenamefont {Balents}\ \emph {et~al.}(2020)\citenamefont
  {Balents}, \citenamefont {Dean}, \citenamefont {Efetov},\ and\ \citenamefont
  {Young}}]{balents-2020-super-stron}%
  \BibitemOpen
  \bibfield  {author} {\bibinfo {author} {\bibfnamefont {L.}~\bibnamefont
  {Balents}}, \bibinfo {author} {\bibfnamefont {C.~R.}\ \bibnamefont {Dean}},
  \bibinfo {author} {\bibfnamefont {D.~K.}\ \bibnamefont {Efetov}},\ and\
  \bibinfo {author} {\bibfnamefont {A.~F.}\ \bibnamefont {Young}},\ }\bibfield
  {title} {\bibinfo {title} {Superconductivity and strong correlations in
  moir{\'e} flat bands},\ }\href {https://doi.org/10.1038/s41567-020-0906-9}
  {\bibfield  {journal} {\bibinfo  {journal} {Nature Physics}\ }\textbf
  {\bibinfo {volume} {16}},\ \bibinfo {pages} {725} (\bibinfo {year}
  {2020})}\BibitemShut {NoStop}%
\end{thebibliography}
\end{document}